\documentclass[12pt]{reportj}
\usepackage{deluxetablej}
\usepackage{hyperref} 
\makeatletter
\def\@to{to}
\makeatother
\usepackage{times}
\usepackage{graphicx}
\usepackage{tocloft}
\usepackage{fancyheadings}
\usepackage{xcolor}
\usepackage{booktabs}
\usepackage{longtable}

\emergencystretch=\maxdimen
\usepackage{datetime}
\newdateformat{ddmonthyyyy}{\THEDAY\ \monthname[\THEMONTH]\ \THEYEAR}

\renewcommand\thesection{\arabic{section}}
\renewcommand\thesubsection{\thesection.\arabic{subsection}}

\def\ssection#1{\setcounter{subsection}{0} \refstepcounter{section} \section*{\hbox to \hsize{\large\bf \arabic{section}. #1\hfill }}\label{sec} \addcontentsline{toc}{section}{\arabic{section}. #1}}
\def\ssubsection#1{\setcounter{subsubsection}{0} \refstepcounter{subsection}\subsection*{\hbox to \hsize{\normalsize\bfseries\itshape \arabic{section}.\arabic{subsection} #1\hfill}}\label{subsec} \addcontentsline{toc}{subsection}{\arabic{section}.\arabic{subsection} #1}}
\def\ssubsubsection#1{\refstepcounter{subsubsection}\subsection*{\hbox to \hsize{\normalsize\it \arabic{section}.\arabic{subsection}.\arabic{subsubsection} #1\hfill}}\label{subsubsec} \addcontentsline{toc}{subsubsection}{\arabic{section}.\arabic{subsection}.\arabic{subsubsection} #1}}

\def\ssectionstar#1{\section*{\hbox to \hsize{\large\bf #1\hfill}} \addcontentsline{toc}{section}{#1}}
\def\ssubsectionstar#1{\subsection*{\hbox to \hsize{\normalsize\bfseries\itshape #1\hfill}} \addcontentsline{toc}{subsection}{#1}}
\def\ssubsubsectionstar#1{\subsection*{\hbox to \hsize{\normalsize\it  #1\hfill}} \addcontentsline{toc}{subsection}{#1}}

\renewcommand{\cftaftertoctitle}{%
\mbox{}\hfill{\normalfont Page}}
\begin{document}

~\\

\vspace{-2.4cm}
\noindent\includegraphics*[width=0.295\linewidth]{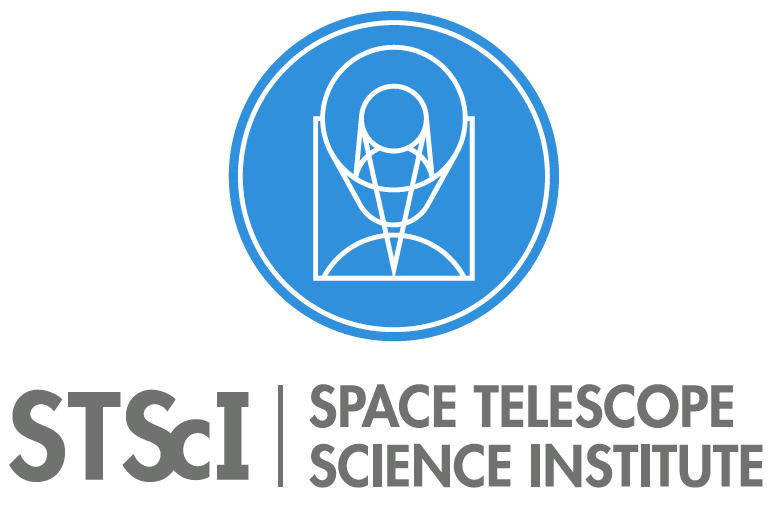}

\vspace{-0.4cm}

\begin{flushright}
{\bf Instrument Science Report COS 2025-18(v1)}
    
\vspace{1.1cm}
    
{\bf\Huge Overview of the New Hubble Spectroscopic Legacy Archive}
    
\rule{0.25\linewidth}{0.5pt}
    
\vspace{0.5cm}
    
Ravi Sankrit$^1$, 
John Debes$^2$,
Matthew Burger$^1$,
Van Dixon$^1$,
Anna Payne$^1$,
Leonardo Dos Santos$^1$,
Thomas Wevers$^{1,3}$,
Travis Fischer$^2$,
Peter Forshay$^1$,
Svea Hernandez$^2$,
Robert Jedrzejewski$^1$,
Rich Kidwell$^1$,
Lauren Miller$^1$,
Marc Rafelski$^1$,
David Rodriguez$^1$,
Robert Swaters$^1$,
Dan Welty$^1$,
Sara Anderson$^1$,
Thomas Bair$^1$,
Joleen Carlberg$^1$,
Brian Charlow$^1$,
Andrew Cortese$^1$,
Tracy Ellis$^1$,
Ben Falk$^1$,
Scott Fleming$^1$,
Elaine Frazer$^1$,
Syed Gilani$^1$,
Alec Hirschauer$^{1,4}$,
Talya Kelley$^1$,
Tim Kimball$^1$,
Jennifer Kotler$^1$,
Adrian Lucy$^1$,
Sunita Malla$^1$,
Christopher Rahmani$^1$,
Fred Romelfanger$^1$,
Kate Rowlands$^2$,
Lisa Sherbert$^1$

\footnotesize{$^1$ Space Telescope Science Institute, Baltimore, MD} \\
\footnotesize{$^2$ AURA for ESA, Space Telescope Science Institute, Baltimore, MD} \\
\footnotesize{$^3$ Schmidt Sciences}
\footnotesize{$^4$ Morgan State University}

\vspace{0.5cm}
    
6 November 2025
\end{flushright}

\vspace{0.1cm}

\noindent\rule{\linewidth}{1.0pt}
\noindent{\bf A{\footnotesize BSTRACT}}

The new Hubble Spectroscopic Legacy Archive (HSLA) provides coadded
spectra of individual targets that have been observed with the
Cosmic Origins Spectrograph (COS) and the Space Telescope Imaging
Spectrograph (STIS) over their operating lifetime.  HSLA uses data
available in the Mikulski Archive for Space Telescopes (MAST). It
automatically produces coadds whenever new data become publicly
available or when there is newly recalibrated data. HSLA defines
individual targets by their associated coordinates, accounting for
proper motions, and uses SIMBAD, NED and the Phase II observing
proposals to obtain astronomical classifications for each object.
Coadded spectra are produced for each observing mode. In the case
of COS far-ultraviolet observations there is one coadded spectrum
for each lifetime position (LP).  Additionally, a spectrum spanning
the entire wavelength range covered by the observations is produced
by abutting the spectra from a selection of individual modes. For
each individual target, HSLA also provides a human-readable metadata
file with key information that can be used in searches or for further
exploration of the data. The HSLA project also makes the code used
for coadding spectra publicly available along with several other
tools (using Jupyter notebooks) for custom coaddition required in
special cases. In this report we will describe the main components
of HSLA and provide a brief description of how the data and metadata
can be accessed.

\vspace{-0.1cm}
\noindent\rule{\linewidth}{1.0pt}


\lhead{}
\rhead{}
\cfoot{\rm {\hspace{-1.9cm} Instrument Science Report COS 2025-18(v1) Page \thepage}}

\renewcommand{\cftaftertoctitle}{\thispagestyle{fancy}}
\tableofcontents


\vspace{-0.3cm}
\ssection{Introduction}\label{sec:intro}

The Hubble Space Telescope's (HST's) two primary spectroscopic
instruments, the Space Telescope Imaging Spectrograph (STIS,
operational since 1997) and the Cosmic Origins Spectrograph (COS,
operational since 2009), have obtained over 64,000 spectra of nearly
7000 individual astronomical objects to date. These instruments
provide access to ultraviolet (UV) wavelengths, with STIS coverage
extending from 1150\,\AA\ to 10,300\,\AA\ covering the far-ultraviolet
(FUV) to the near-Infrared (NIR), and COS coverage from 912\,\AA\
to 3300\,\AA\ spanning the Lyman-ultraviolet (LUV), FUV and the
near-ultraviolet (NUV) regimes.

The wealth of UV spectra from COS and STIS are hosted by the Mikulski
Archive for Space Telescopes (MAST) and are available to the public,
but they are primarily organized by links to individual observing
programs in which they were obtained. Therefore finding and selecting
data for individual objects or by source type for several objects
is not straightforward. The need to organize the spectra in a simpler
way was recognized years ago, and resulted in the creation of the
original Hubble Spectroscopic Legacy Archive (o-HSLA; Peeples et
al.\ 2017). The o-HSLA represented a significant step forward in
the organization of HST UV spectra and provided a template for
organizing a legacy archive. However, the data products were static
and did not include STIS observations.  Only COS NUV and FUV data
obtained before July 2018 are in the o-HSLA, and they have not been
updated with any newer data products or with improved calibrations.

The new Hubble Spectroscopic Legacy Archive (HSLA) was motivated
by the same underlying reasons as the earlier version, but it was
designed from the start to include spectra from all COS and STIS
channels and to automate the process to keep the archive current.
One of the drivers leading to the implementation of the new HSLA
was the recently completed HST Director's Discretionary program,
Hubble UV Legacy Library of Young Stars as Essential Standards
(ULLYSES). ULLYSES provided the impetus for creating coaddition
software for STIS and COS spectra and providing scientifically
useful data products that could be easily accessed by the astronomical
community (Roman-Duval et al.\ 2025). The data in the ULLYSES program
are relatively constrained both in terms of target type (young hot
stars) and in the observation modes used, but it was realized that
the software tools were sufficiently flexible that they could be
adapted more generally.

The HSLA was conceived as a multi-stage project. The first stage
was to provide coadditions of all spectra of each target obtained
as part of a single observing program at the visit and program
level, sorting them by instrument and grating.  Additionally, at
each level abutted spectra combining instruments and modes are
provided.  These show the full wavelength coverage of the data, and
are meant for ``quick-look'' purposes rather than for use in
scientific analysis.  The independent value of these visit and
program level data products was realised early during implementation,
and this first stage was completed as the Hubble Advanced Spectral
Products (HASP) project (Debes et al.\ 2024).

This document describes the completion of the second stage and the
release of the full HSLA, in which coadded spectra for each individual
target across multiple programs are created. These are made available
along with target classifications and metadata files and enable
archive-wide searches based on individual source names, or broader
categories of source types. The following sections describe the key
components of the HSLA: target association (\S2), target classification
(\S3), and the data products (\S4).  Following that, in \S5, we
describe the testing procedure used to verify and validate the HSLA
products.  Information on catalog and data access as well as a list
of cases requiring special consderation are given in \S6, and \S7
describes python scripts (implemented as Jupyter notebooks) that
have been written to perform custom coadditions of the data.

\vspace{-0.3cm}
\ssection{Target Association}\label{sec:assoc}

The HST archive contains over 60,000 COS and STIS exposures distributed
over a few thousand observing programs. The main tasks of target
association are to identify all the unique sources observed, to
determine all the exposures (and thereby calibrated data files)
associated with each source, and to label each source with a unique
name and provide their best possible coordinates.

In principle, the main task of cross-matching is straightforward,
as one can use target coordinates and names to make the identifications.
In practice, there are a number of complicating factors. The primary
sources of information are the names in the submitted Phase II
proposals and the actual coordinates observed. It is often the case
that the same source targeted in different programs will be identified
using different names, and may have slightly different coordinates.
There are several reasons for scatter in the coordinates for the
same target. In the case of objects where proper motion data are
provided, the header keywords in the calibrated data files are
estimated at the most likely execution time. Other sources of
potential uncertainty include the fact that different instrument/grating
combinations are located at slightly different sky positions, and
the possibility of incorrect specification of the reference frame
by the user.

The outputs of the target association step are used for target
classification (\S3) and for creating the coadded data products
(\S4), which include a human-readable metadata file for each target.

\subsection{Optimal angular radius for cross-matching} 

The first task for target association is finding the optimal angular
tolerance for identifying unique targets based on sky coordinates.
This is non-trivial because several observations have been made in
crowded fields (such as the Magellanic Clouds) where the separation
between individual sources can be smaller than the coordinate scatter
for individual targets. The cross-matching radius should be large
enough such that it encompasses the scatter, but not so large that
it is unable to distinguish between multiple nearby sources. Using
a very small cross-match radius will improve accuracy at the cost
of having multiple products for a unique target, while a large
cross-match radius will improve completeness at the cost of coadding
unrelated exposures. For a given angular tolerance for matching,
the accuracy can be improved by using the names for targets provided
by the user in the Phase II proposal or from SIMBAD\@.  To determine
the optimal cross-matching radius for target association, we first
performed a name-based cross-match of every individual exposure
with every COS+STIS observation in the HSLA archive (excluding
moving targets, which are so specified in the Phase II proposals,
and those labeled as ``calibration'' observations).  For every
exposure we included the full list of SIMBAD identifiers, thereby
generating a list of exposures that is potentially associated with
every observation. For every exposure, we can examine the distance
distribution of every potentially associated observation based on
target name. This is useful to understand the typical scatter in
the spatial distribution of exposures that are highly likely to be
associated based on their names, and it allows us to quantify the
success (or lack thereof) of name-based target association as a
function of on-sky distance.

To test this in practice, we need to have a ground truth value of
the object coordinates available. We treated the SIMBAD coordinates
(which are typically those provided in Gaia DR3) as the actual
source position, and performed a cross-match (where the cross-match
radius is the parameter to be optimized) with the positions listed
in the headers of the individual exposure calibrated data files.
We define the {\it accuracy} as the number of (SIMBAD) sources for
which every associated file (based on name matching only) is located
within the cross-match radius under consideration. The completeness
as a function of cross-match radius is derived from the number of
SIMBAD sources with associated exposures; for example, if we consider
a 0.5$^{\prime\prime}$ cross-match radius and every associated file
is located at a distance $>$0.5$^{\prime\prime}$ from the SIMBAD
position, there will be no associations. Finally, we quantify the
contamination as the number of sources where exposures are present
within the cross-match radius but are not associated through their
name. This happens when multiple sources with different names are
located close together (such as individual stars in a dense cluster)
and would result in exposures erroneously coadded if associated.

\begin{figure}
\centering
\includegraphics[width=0.8\linewidth]{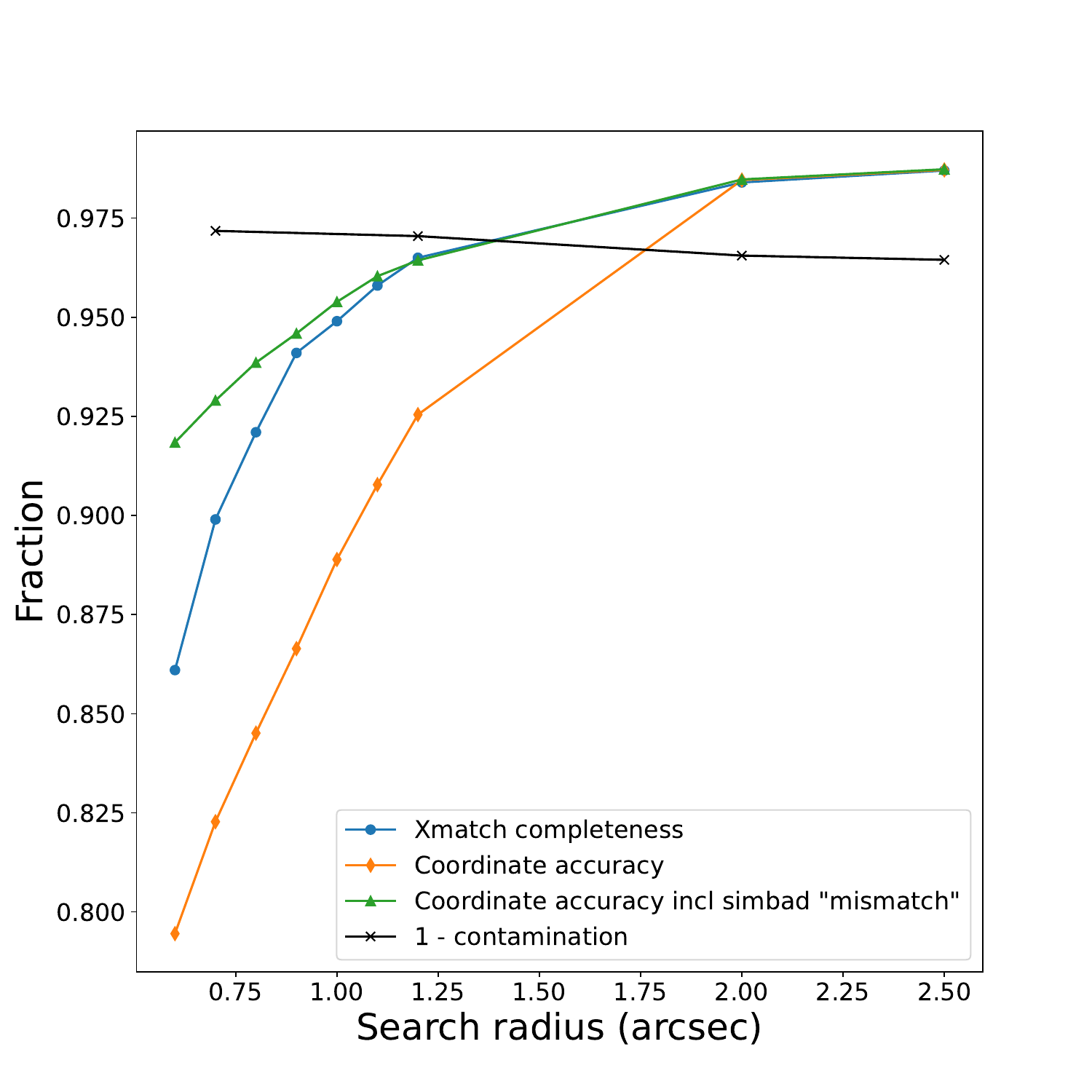}
\caption{Completeness, accuracy and contamination of name-based
target association as a function of cross-match radius. The
cross-match completeness is defined as the number of apertures
that are identified as cross-match divided by the total number
of apertures that match one of the target SIMBAD aliases as a
function of radius. Coordinate accuracy is defined as the number
of apertures that, based on name and coordinates, are correctly
cross-matched to a given target as a function of radius. A
``mismatch'' corresponds to a SIMBAD cross-match based on name,
but the difference between SIMBAD coordinates and the HST
observation is $>2^{\prime\prime}$.}
\label{fig:xmatch}
\end{figure}

We tested a range of cross-match radii from 0.3--2.5$^{\prime\prime}$,
and considered the trade-off between maximizing the completeness
while simultaneously maintaining high accuracy. Figure \ref{fig:xmatch}
shows these quantities as a function of cross-match radius. The
results are very encouraging, as a completeness and accuracy of
$>$95\% can be achieved at a cross-match radius of 2$^{\prime\prime}$,
while the contamination remains of order a few percent (the black
curve shows the uncontaminated fraction). We therefore choose
2$^{\prime\prime}$ as an optimal cross-match radius for target
association.

\subsection{Procedure details}

With the optimal cross-matching angular radius determined, the
target association is done using coordinate matching. Additionally,
it is assumed (as was done for HASP) that within a given program
there is a one-to-one matching between unique targets and Phase II
names.

Initially, a list was made of all the COS and STIS datasets used
in HASP visit and program level products along with the relevant
metadata: right ascension (RA), declination (DEC), observation time,
proper motion and observation epoch, which are stored in the archive
catalog. Then, the proper motions were used to determine the
coordinates at a common epoch (January 1, 2016) for all the datasets.
The observations were then grouped by visit and target name supplied
in the Phase II proposals, with the unique visit names determined
by the first six characters in the dataset root names. This resulted
in a list of visit/target combinations, each of which is called a
``pointing''. The next step was to loop through the list of pointings
and determine the angular separation of each pointing from all other
pointings (using the astropy SkyCoords routine). All observations
within 2$^{\prime\prime}$ of the pointing were considered to be
part of the same association. If no previous targets were found
within 2$^{\prime\prime}$ of a given pointing, it was designated
as a new target; otherwise it was added to the existing target.
The RA and DEC of the target were taken to be the mean of the RAs
and DECs of all the associated observations, and when new observations
were added to an existing target, the coordinates were updated.

As the RA and DEC of each target could change during the process
of assembling the associations, a check was made after all the
datasets had been associated with a target to ensure that all the
coordinates were within 2$^{\prime\prime}$ of the target center.
When this was not the case, the observations furthest away from
each other (extreme pointings) were identified, and each of the
remaining pointings were associated with the extreme pointing closest
to them.  This is an issue only in crowded fields where there are
observations of distinct objects within a few arcseconds of each
other. A special, possibly unique, case to be noted is the pair of
targets, $\alpha$\,Cen~A and B, which move relative to one another
non-linearly in a way that causes coordinate confusion.

After associating targets, a list was compiled of all the target
names, target categories and target descriptions supplied in the
Phase II proposals for each of the constituent datasets, which are
stored in the headers (keywords: TARGNAME, TARGCAT, TARGDESCR,
respectively) of the input FITS files. This list is not a primary
HSLA data product but is used to help with checking and testing the
classification and categorization of targets. The targets thus
defined will remain static in HSLA operations. New targets will be
added to HSLA as they are observed, but when new datasets are added
to existing target associations, the coordinates will not be updated.

\vspace{-0.3cm}
\ssection{Target Classification}\label{sec:class}

There are a wide range of target types in the HSLA, and a robust
classification scheme is essential for efficient search and retrieval
of data on objects in a specific class of scientific interest, such
as ``White Dwarf Stars'' or ``Ultraluminous Infrared Galaxies''.
One part of the problem is deciding what the classes should be, or
in other words defining a vocabulary for the classification, and
the other is to decide which class or classes a given object should
be assigned to. In the o-HSLA, the classification was done by
manually sorting indivdual targets into tables for several broad
categories, which are described in Peeples et al.\ (2017).  These
categories and tables, like the rest of the o-HSLA, are static.
For the new HSLA the primary requirement was to automate assigning
target classifications to existing and new data, and an important
consideration was to make the classification scheme maximally useful
for archival research.

\subsection{Methodology}

HSLA's target classification is structured as a three-tiered
hierarchy. This is to faciliate both broad and narrow searches.
For example, a user can query all O-type stars, or all Early Type
stars, or even all stars. This approach offers users a flexible
method for interacting with the database, making it suitable for a
wide variety of research use cases.

We used three sources of information, SIMBAD, NED and the
Phase II proposals to develop a database of all possible object
types and create the classification vocabulary for HSLA\@.  SIMBAD
is a rich resource for a wide variety of astronomical objects from
stars to nebulae to external galaxies. NED is an extragalactic
database, and therefore more specialized and restricted in scope
than SIMBAD\@. In the case of the Phase II proposals, the principal
investigators (PIs) are required to include scientific keywords
describing their targets. These have the advantage of both flexibility
and the specialized knowledge of the proposing team, but this
flexibility also allows for greater inconsistencies, particularly
when classifications are based on the scientific aim of the proposal
rather than describing the observed target.

The available classifications in all three sources were scrutinized
carefully, and a broad amalgamation of categories were defined in
three hierarchical tiers, with Tier 1 providing the most general
description, and Tier 2 and 3 increasing in specificity. This
three-tiered structure is the basis for HSLA target classification
and enables searches on broad and narrow categories of objects.  

\begin{figure}
\centering
\includegraphics[width=\linewidth]{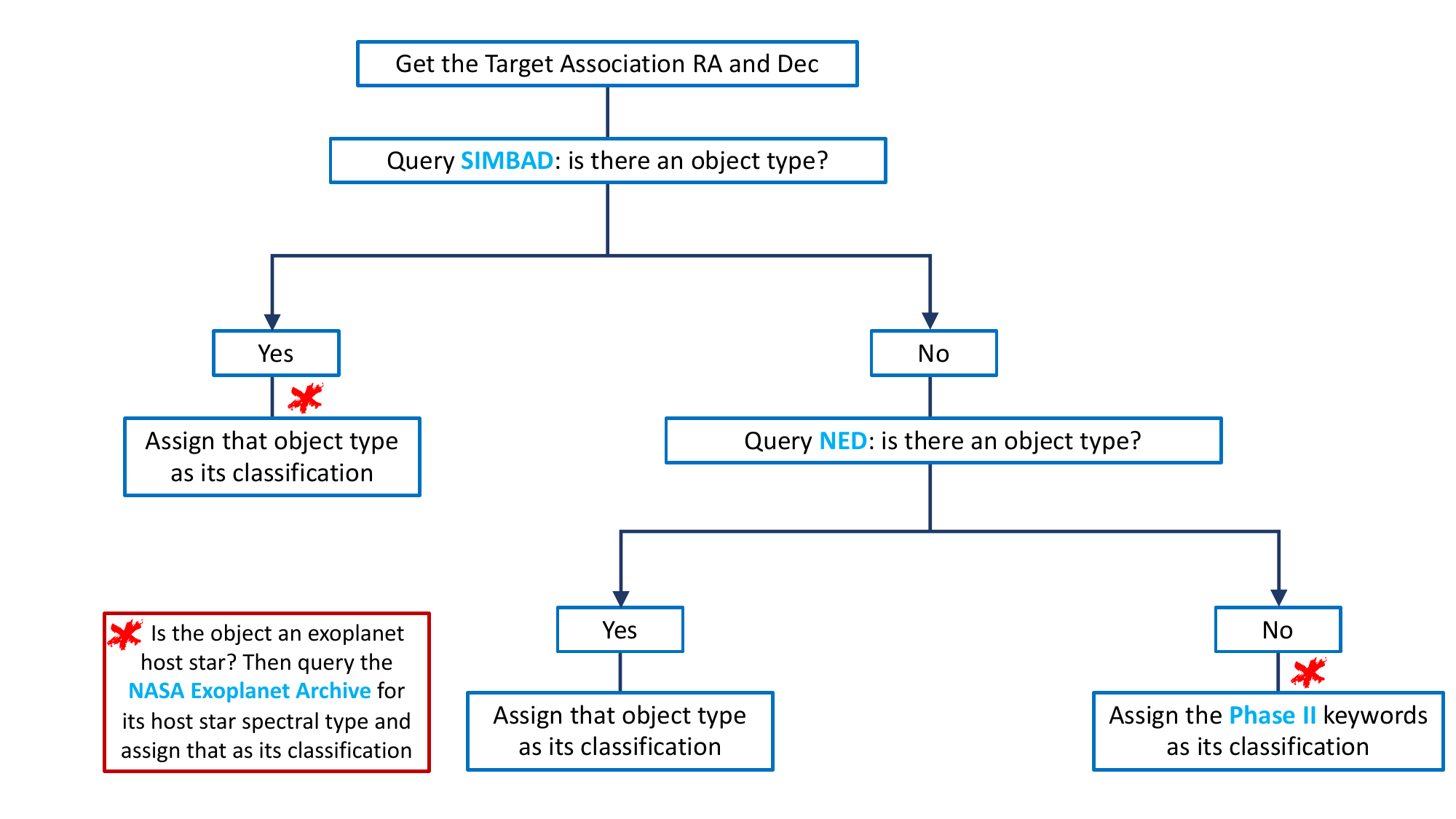}
\caption{Flow chart detailing the decision tree of the final object
classification. The databases SIMBAD, NED, and the NASA Exoplanet
Archive are queried and assigned as the classification in that order
of object type availability, but if the target is not present in
any, then the Phase II keywords are assigned as the classification.
The red asterisks denote the location in the decision tree when an
additional query is performed in the case the object is an exoplanet
host star. The NASA Exoplanet Archive is queried for the host star
spectral type, which is then assigned as its object type.}
\label{fig:decisiontree}
\end{figure}

For every target in the HSLA database, SIMBAD, NED and the Phase~II
proposals are queried in that order, and the object types available
are assigned to the target. If an object type exists in SIMBAD for
the target in question, those classifications are used for HSLA. If
there is no object type, then the description returned by NED is
used if available. If the target is present neither in SIMBAD nor
in NED, then the Phase II keywords are used. A special case is if
the Phase II classification indicates an exoplanet, which triggers
a query to the NASA Exoplanet Archive. The HSLA decision tree in
how an object type is assigned is graphically represented as a flow
chart in Figure \ref{fig:decisiontree}. 

The three main Tier 1 classes are ``Galaxy'', ``ISM'' and ``Star''.
There are 12 Tier 2 and 59 Tier 3 classes under ``Galaxy'', 14 and
46 under ``ISM'', and 21 and 140 under ``Star''. In each of the
secondary and tertiary classification lists, one of the options is
``Pass'', which is assigned if there is insufficient information
about the target to classify it at these narrower levels.  For some
targets neither SIMBAD nor NED types are available, and the Phase~II
keywords make the classification ambiguous. These are assigned a
Tier 1 classification of ``Multiple''.  A handful of observations
do not fit into any of the categories and are classified as
``Unknown''. A potential future plan is to manually examine the
targets falling in these latter two categories and reassigning them
to the most appropriate classes.  A hierarchical list of all the
classes is available on the
\href{https://archive.stsci.edu/missions-and-data/hst/hsla/classifications}
{HSLA website}.

In addition to the three-tiered classification vocabulary, each
Tier 1, 2 and 3 combination was mapped to a Unified Astronomy
Thesaurus (UAT) concept and number (Frey \& Accomazzi 2018).  In
most cases the UAT concepts match the Tier 3 classifications -- for
example HSLA Tier 3 ``Seyfert galaxy'' maps to UAT concept (ID)
``Seyfert galaxies (1447)''. In other cases, they are closely related
-- ``Blue supergiant'' in HSLA maps to ``Early-type supergiant stars
(431)''. The \href{https://astrothesaurus.org/}{UAT project} provides
brief definitions of all the object types as well as links to related
types. Mapping classes to UAT allows HSLA users to explore other
archives and documents based on a wider vocabulary and to access
supplementary data or information on specific types of astronomical
objects of interest.

\subsection{Results}

\begin{deluxetable}{lr}
\tabletypesize{\small}
\tablewidth{0pt}

\tablecaption{Number of HSLA Targets in various Classes
\label{tab:hsla_class_totals}}
\tablehead{\colhead{Classification} & \colhead{Target Count}}
\tablecolumns{2}
\startdata
\sidehead{Tier 1}
Galaxy                 & 2459 \\
ISM                    &  179 \\
Star                   & 3894 \\
Multiple               & 170 \\
Unknown                &  10 \\

\midrule

\sidehead{Tier 2}
Active galaxy          & 2082 \\
Dwarf galaxy           &  158 \\
Nebulae                &  145 \\
Stellar Remnant        & 1004 \\
Early-type star        &  889 \\
Late-type star         &  766 \\
Binary star            &  487 \\
Young stellar object   &  250 \\

\midrule

\sidehead{Tier 3}
Quasar                 & 963 \\
Seyfert galaxy         & 707 \\
Starburst              & 193 \\
Planetary Nebula       &  64 \\
White dwarf            & 870 \\
B star                 & 255 \\
Cataclysmic variable star & 186 \\
Blue supergiant        & 166 \\

\enddata
\end{deluxetable}

The results in this section are based on the HSLA run implemented
on September 8, 2025, which used data collected through June 25,
2025.  A total of 6712 unique HSLA targets were identified.  Of
these, an overwhelming majority have a Tier 1 classification of
``Star'' (3894) or ``Galaxy'' (2459).  Only a small fraction (179)
fall under ``ISM''.  Of the remaining, 170 are classified as
``Multiple'', and 10 as ``Unknown''.

The number of targets classified in each of the main Tier 1 classes,
along with the numbers for a sample of Tier 2 and Tier 3 classes
are shown in Table \ref{tab:hsla_class_totals}. It is worth noting
that a large fraction of the Galaxies have ``Active galaxy'' and
``Quasar'' as their secondary and tertiary classifications, and
that among the Stars, the largest type are ``Stellar Remnants''
(secondary), which include ``White dwarf'' (tertiary). Of the 6712
total HSLA targets, SIMBAD was used to classify 5548, of which 58
were exoplanet host stars requiring NExScI\@. NED was used for 38
objects, and the remaining 1126 used the Phase~II proposals for
classification.

\subsection{Summary}

HSLA is providing automated target classifications for all the
targets with COS or STIS spectral observations. The classification
scheme uses three hierarchical tiers, and enables searches based
on object type at all three levels, e.g. from ``Stars'', to
``Early-type star'' to ``O star'', or ``Galaxy'' to ``Active galaxy''
to ``Starburst''. The classifications at each tier were developed
based on a careful scrutiny of the object descriptions in SIMBAD,
NED and the Phase II proposals. The process of assigning categories
is automated and will be applied to all new targets observed as
part of the HSLA process. The scheme and implementation are
sufficiently flexible and the vocabulary is sufficiently comprehensive
that they may be used for high-level data product archives from
other general purpose instruments.

\vspace{-0.3cm}
\ssection{Data Products}\label{sec:datap}

Once a selection of one-dimensional spectra are successfully
associated with a target, coaddition occurs following the same
procedure as described for HASP (Debes et al., 2024). Input spectra
are drawn from the pre-filtered datasets identified for HASP,
followed by a coaddition of all input spectra from multiple CENWAVES
or apertures within a single grating across one or more programs
that have targeted the same object.

The largest source of potential errors comes from observations in
which the source is not well centered or has drifted out of the
slit/aperture, and the quality issues have gone unreported. To avoid
these errors, spectra with anomalously low flux are excluded from
the coadds.  As with HASP data, this procedure makes default HSLA
data products inaccurate for variable sources (Debes et al., 2024),
but custom coaddition notebooks in the
\hyperlink{https://spacetelescope.github.io/hst_notebooks/notebooks/HASP/FluxScaleTutorial/FluxScaleTutorial.html}{HASP
Jupyter notebook repository} provide guidance about how to create
a useful coadd from flux variable spectra. No flux checking is
performed across gratings.  Observations of extended sources that
have spectra with different aperture sizes may have flux offsets
between gratings.  As a reminder, our flux and wavelength requirements
do not apply to extended sources. Additionally, variable sources
that vary significantly between epochs may show flux offsets between
gratings.

The coadd process produces data products across multiple HST programs
for each grating, each lifetime position (LP) for COS FUV modes,
and a target quicklook spectrum that covers the full wavelength range
of the available observations. For well-studied objects this can
span from $\sim$900-10,000~\AA, covering the far-UV to the near-IR.

The data products are in FITS table format with two extensions. 
The first extension includes a wavelength column in units of
Angstroms, flux and error columns in units of
erg~s$^{-1}$~cm$^{-2}$~\AA$^{-1}$, a Signal-to-Noise Ratio (SNR)
column, and an effective exposure time column in units of seconds.
The wavelength bin size for single grating spectra is set by looking
at the wavelength differences between adjacent bins of all the input
spectra, and setting the product bin size to the largest of these
differences.  The wavelength bins are of equal width in these
spectra.  The quicklook spectra inherit the wavelength bins from
the constituent single-grating spectra, and therefore the bin widths
will generally change with the transition from one single-grating
spectrum to the next that make up the abutted spectrum.  The second
extension is the provenance table, which is described in more detail
in Section \ref{subsec:provenance}.

\subsection{Coadd Naming Conventions}\label{subsec:dpnaming}
The naming convention for data products is different for single
grating products and for the quicklook spectra:

\noindent Single Grating: \\ {\small{\texttt{hst$\_$<instrument>$\_$<target>$\_$<opt$\_$elem>$-$<lp\#>$\_$cspec.fits}}} \\

\noindent Target Quicklook: \\ {\small{\texttt{ hst$\_$<target>$\_$aspec.fits}}} \\

\noindent In the above naming scheme, \texttt{instrument} is either COS or
STIS, \texttt{target} refers to the target association name, and
\texttt{opt$\_$elem} is the grating name for the relevant mode(s).
If a grating name is shared by COS and STIS a ``c'' or ``s'' will
denote which instrument it is associated with.  Observations with
COS gratings can be obtained at a variety of lifetime
positions\footnote{https://hst-docs.stsci.edu/cosdhb/appendix-a-cos-lifetime-positions/a-1-cos-lifetime-positions}.
The lifetime position is denoted in the filename with {\texttt{lp\#}}.

\subsection{Provenance Extension}\label{subsec:provenance}

Both HASP and HSLA products contain a provenance table that contains
useful information about the input filenames, proposal IDs, and
instrument details. Table \ref{tab:provenance} gives a detailed
breakdown of what information is included. The \texttt{coadd} code
reports all input data files in the provenance table regardless of
whether some input files are removed due to certain conditions as
described in Debes et al., (2024). A careful parsing of both the
provenance file and the output trailer file described in Section
\ref{subsec:trailer} can help determine what input files were
combined in the the different data products. Some examples of how
to parse HASP trailer files, which will have similar formatting to
HSLA trailer files, are available as a
\href{https://spacetelescope.github.io/hst_notebooks/notebooks/HASP/DataDiagnostic/DataDiagnostics.html}{Jupyter
notebook}.

\begin{deluxetable}{clll}
\tabletypesize{\small}
\tablewidth{0pt}

\tablecaption{Provenance Table Columns
\label{tab:provenance}}
\tablehead{\colhead{Field} & \colhead{Name} & \colhead{Units} & \colhead{Description}}
\startdata
1 &  FILENAME  &  & \\        
2 &   EXPNAME  &  & Dataset Name \\      
3 &  PROPOSID  &  & Program ID \\        
4 & TELESCOPE  &  & \\        
5 & INSTRUMENT &  & \\         
6  & DETECTOR  &  & \\         
7  & DISPERSER  &  & Grating Used \\      
8  &  CENWAVE  &  & \\         
9  & APERTURE  &  & Slit Used \\       
10  & LIFE\_ADJ  & & Detector Lifetime Position \\       
11  &  SPECRES  &  & Spectral Resolution \\       
12  &  CAL\_VER  &   & Pipeline Version \\      
13  &  MJD\_BEG  & d & Exposure Start \\       
14  &  MJD\_MID  & d & Exposure Midpoint \\      
15  &  MJD\_END  & d & Exposure End \\     
16  &  XPOSURE  &  s & Exposure Time \\      
17  &  MINWAVE & \AA\ & Minimum Wavelength \\     
18  &  MAXWAVE & \AA\ & Maximum Wavelength \\        
 \enddata
\end{deluxetable}

\subsection{Coadd abutment prioritization} \label{subsec:dpprioritization}
Different gratings within the full wavelength quicklook spectrum
are abutted. Abutment is when two or more gratings are joined at a
single transition wavelength, rather than merged (for more details
on how abutment works for HASP and HSLA, see Debes et al., 2024).
Abutment is used in both HASP and HSLA products due to the varying
resolutions of the various modes and the difficulty involved in
coadding overlapping regions. The transition wavelengths and the
priority of a given grating is provided by a separate JSON
prioritization table.

In some cases, a target has been observed at the same wavelength
by multiple gratings. In these situations, only a subset of gratings
populate the final quicklook spectrum. The motivation for which
gratings are prioritized is based on optimizing line detection for
a median observation within a given grating.  The line detection
sensitivity of a grating, `gk', can be estimated based on the
limiting equivalent width of an observation ($\mathrm{EQW_{limit}}$)
for an unresolved spectral line:

\begin{equation}
\label{eqn:limit}
\mathrm{EQW_{limit}}(\lambda) \propto \sigma_{\mathrm{cont}} \frac{\lambda}{R_{\mathrm{gk}}}
\end{equation}

\noindent where $\sigma_{\mathrm{cont}}$ is the uncertainty per
resolution element, $\lambda$ is the wavelength, and $R_{gk}$ is
the resolution of the grating. For spectra with heterogeneous
SNR, it is preferable to recast Equation \ref{eqn:limit} in terms
of instrument sensitivity($S_{\mathrm{{gk}}}$), the exposure time
($t_{\mathrm{exp,gk}}$), and the source flux ($F(\lambda)$):

\begin{eqnarray}
 \mathrm{EQW_{limit,g1}}(\lambda) = & \left[S_{\mathrm{g1}}(\lambda)t_{\mathrm{exp,g1}}F(\lambda) \right]^{-1/2} {\frac{\lambda}{R_{\mathrm{g1}}}} \\
 \mathrm{EQW_{limit,g2}}(\lambda) = & \left[S_{\mathrm{g2}}(\lambda)t_{\mathrm{exp,g2}}F(\lambda) \right]^{-1/2} \frac{\lambda}{R_{\mathrm{g2}}}
\end{eqnarray}

\noindent where $S_{\mathrm{g1}}(\lambda)$ and $S_{\mathrm{g2}}(\lambda)$
are the wavelength dependent sensitivities of gratings g1 and g2,
respectively, and $t_{\mathrm{exp,g1}}$ and $t_{\mathrm{exp,g2}}$
the median exposure time across all observations obtained with these
gratings.

Grating g2 will have higher priority if the following inequality is satisfied:
\begin{equation}
\left[S_{\mathrm{g2}}(\lambda)t_{\mathrm{exp,g2}}R_{\mathrm{g2}}^2 \right]^{-1/2}  < \left[S_{\mathrm{g1}}(\lambda)t_{\mathrm{exp,g1}} R_{\mathrm{g1}}^2\right]^{-1/2}
\end{equation}

A table of median exposure times for each grating was generated
from archived COS and STIS data, and used in equation (4) to determine
the prioritization. These are listed in Table \ref{tab:priorities}
for HSLA products. The publicly available {\texttt{coadd}} code
allows for the customization of prioritization for specific science
cases.  In particular, the overlapping regions of COS/G130M and
COS/G160M may have line sensitivities that depart from the simple
formulation above and require special handling (Section \ref{sec:custom}).

\begin{deluxetable}{cccc}
    \tabletypesize{\small}
    \tablewidth{0pt}
    \tablecaption{
    Grating Abutment Order Priorities (Highest to Lowest Priority)
    \label{tab:priorities}} 

\tablehead{
    \colhead{Instrument}&\colhead{Grating}&\colhead{Wavelength Range (\AA)}&\colhead{\#CENWAVEs/Tilts}}
\startdata
STIS Echelle & E140H & 1140--1700 & 3 \\
STIS Echelle & E230H & 1620--3150 & 6 \\
STIS Echelle & E140M & 1144--1710 & 1 \\
COS FUV & G130M & 900--1470 & 8 \\
COS FUV & G160M & 1342--1800 & 6 \\
STIS Echelle & E230M & 1605--3110 & 2 \\
COS NUV & G185M & 1664--2134 & 15 \\
COS NUV & G285M & 2474--3221 & 17 \\
COS NUV & G225M & 2069--2526 & 13 \\
STIS MAMA & G140M & 1140--1740 & 12 \\
STIS MAMA & G230M & 1640--3100 & 18 \\
STIS CCD & G430M & 3020--5610 & 10 \\
STIS CCD & G750M & 5450--10140 & 9 \\
COS FUV & G140L & 900--2150 & 3 \\
STIS CCD & G230MB & 1640--3190 & 11 \\
STIS MAMA & G140L & 1150--1730 & 1 \\
STIS MAMA & G230L & 1570--3180 & 1 \\
STIS CCD & G430L & 2900--5700 & 1 \\
STIS CCD & G230LB & 1680--3060 & 1 \\
STIS CCD & G750L & 5240--10270 & 1 \\
COS NUV & G230L & 1650--3200 & 4 \\
\enddata
\end{deluxetable}

\subsection{Metadata Files}\label{subsec:dpmetadata}

In addition to the spectra, each HSLA target association has a
companion metadata text file that compiles information about the
target and the input data in a simple human-readable format. We
present two examples and describe the contents of these files.

\begin{figure}
\centering
\includegraphics[width=0.9\textwidth]{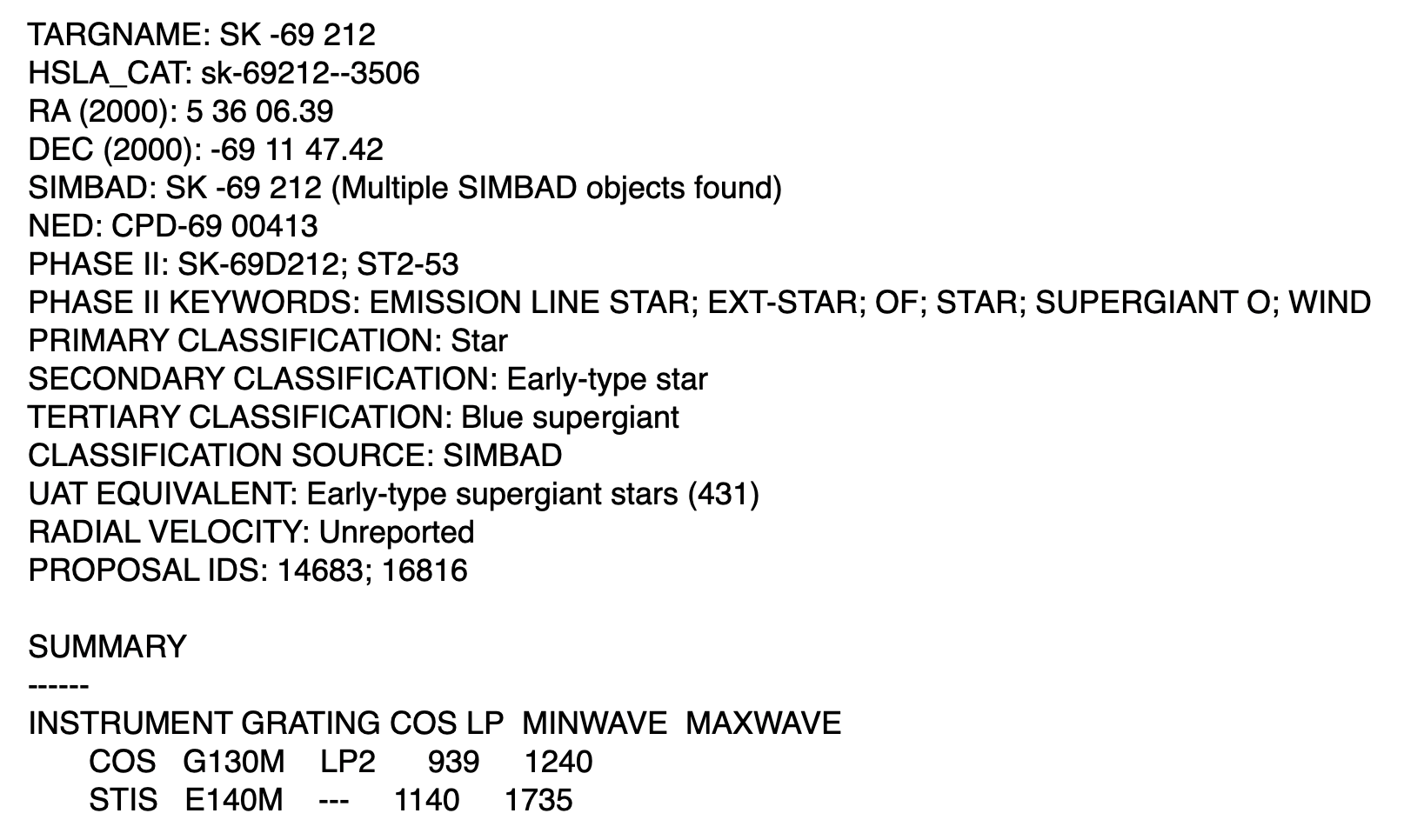}
\caption{Example metadata file for the LMC O star SK~-69~212.}
\label{fig:sk69212metadata}
\end{figure}

\begin{figure}
\centering
\includegraphics[width=0.95\textwidth]{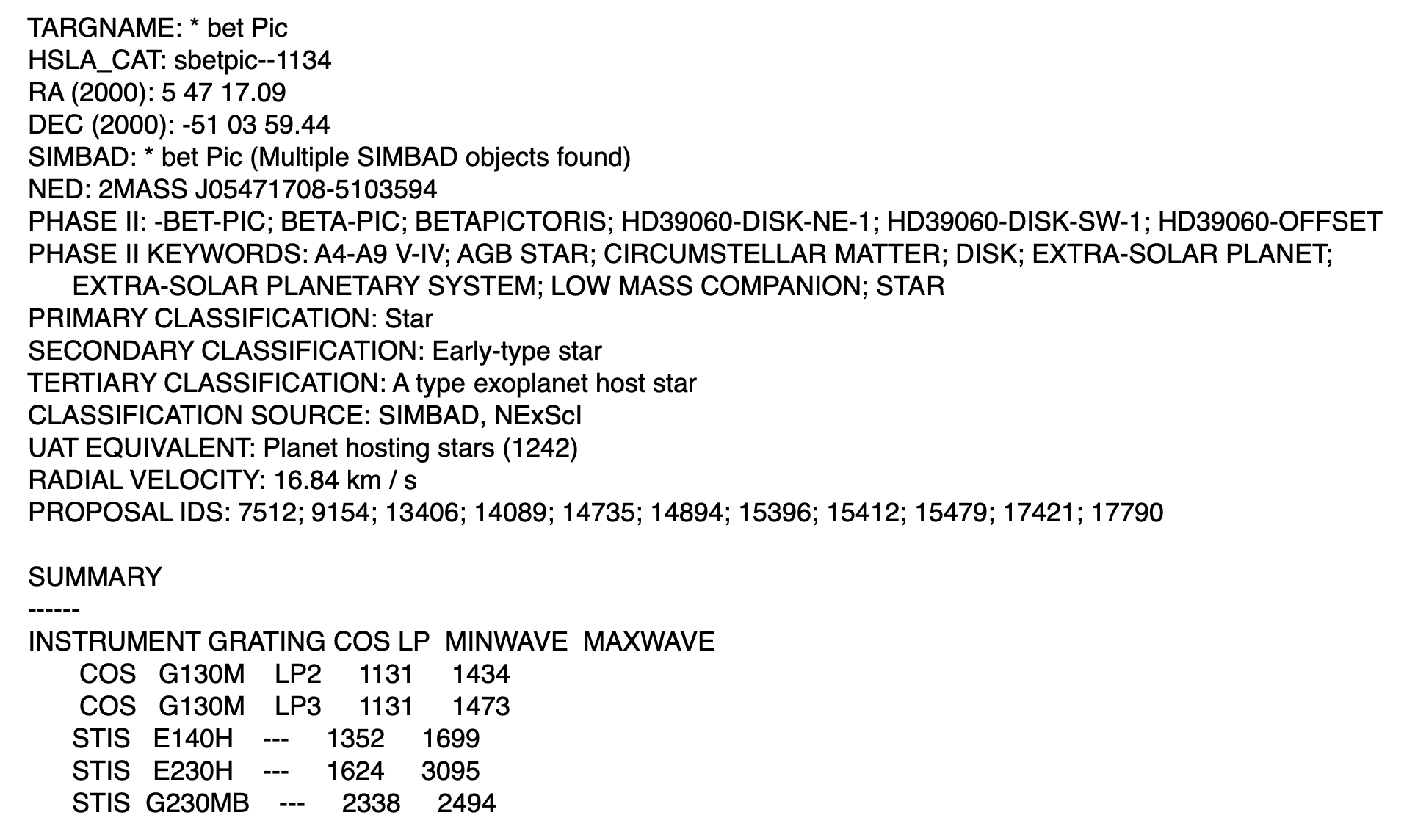}
\caption{Example metadata file for the exoplanet hosting A star
$\beta$-Pictoris.}
\label{fig:betapicmetadata}
\end{figure}

The first is the file for SK~-69~212, an O star in the Large
Magellanic Cloud (LMC; Fig.~\ref{fig:sk69212metadata}). The first
part of the file contains naming and coordinate information.  A
primary target name and set of coordinates is derived from the
target association process described in Section \ref{sec:assoc}.
J2000 coordinates at epoch 2016 are derived from either SIMBAD,
NED, or the Phase II astrometry.  Next, there is a unique HSLA
catalog name with the primary target name plus a unique numerical
identifier, along with names, if available, from SIMBAD, NED, and
the Phase II naming. If SIMBAD reports other sources within
2$^{\prime\prime}$ of the target, this is noted along with the name.
This is useful for checking for possible contaminants. In this case
there is an X-ray source identified by Chandra about 0.3$^{\prime\prime}$
away, which does not impact the HSLA catalog nor the data. The next
few entries deal with the target class, including each of the three
tiers of classification, the primary classification source and the
UAT equivalent classification.  Following that is a line reporting
either the radial velocity or the redshift (for external galaxies)
if available in SIMBAD, which it is not for SK~-69~212.  The last
part of the metadata file lists the proposal IDs from which the
data were derived, followed by a summary of the constituent modes
along with the wavelength coverage, and LP information for COS FUV
observations.

The second example is the well-studied A star, $\beta$-Pictoris,
which hosts exoplanets (Fig.~\ref{fig:betapicmetadata}). The metadata
file shows at a glance that this is a popular target, having been
observed in 11 programs and with several COS and STIS modes. As it
is an exoplanet host, NExScI was queried to obtain the classifcation
in addition to SIMBAD (see Fig.~\ref{fig:decisiontree}), and this
information is included in the Tertiary Classification as well as
the UAT Equivalent fields. The target also has a reported radial
velocity in SIMBAD, which is included in the metadata file. The use
of HSLA data for studying $\beta$-Pictoris is discussed at more
length in \S7.2, below.

\subsection{Trailer Files} \label{subsec:trailer}

In addition to the metadata file and the coadds that are delivered
with an HSLA target, users can also download a trailer file (denoted
with the ``.trl'' file extension) that contains all the text output
from the \texttt{coadd} code, in a format similar to HASP products.
The outputs are particularly useful for determining whether particular
datasets were included or rejected during the coaddition process
by searching for particular key words. Details of how to inspect
trailer files for dataset rejection are described in the HASP Jupyter
notebook on
\href{https://spacetelescope.github.io/hst_notebooks/notebooks/HASP/DataDiagnostic/DataDiagnostics.html}{Data
Diagnostics} and with some modification can be applied to HSLA
trailer files when combined with information from the metadata file.

\vspace{-0.3cm}
\ssection{Testing}\label{sec:test}

We conducted several tests to ensure the robustness of the HSLA
products. First, a random sample of 100 HSLA targets was inspected
in detail to verify the accuracy of their coordinates, the accuracy
and completeness of the target associations, and the accuracy of
their classifications. Then, following the procedure used for HASP,
we tested the agreement in flux and wavelength between input spectra
and the output coadds for each HSLA target with a requirement that
$>$75\% of spectra matched to within 5\% in flux and within half a
resolution element in wavelength. Finally, we used a subset of the
CALSPEC standard stars to verify the absolute flux calibration
accuracy of the HSLA coadded spectra.

\subsection{Target Association and Classification Accuracy}

Investigating over 6000 individual targets for accuracy of
classifications, naming, coordinates and association accuracy
reliably would be prohibitively time intensive.  Therefore we instead
conducted random sampling of the HSLA targets. We set our accuracy
requirements to be consistent with $>$90\% for target association
completeness and accuracy, and a best effort with respect to automated
classifications.

A random sample of 100 targets was selected for detailed examination.
This sample size ensured that under the assumption of binomial
statistics, we could demonstrate that we had met our requirements
provided the number of failures were small and under the assumption
of 95\% confidence when calculating a cumulative probability
distribution.

Our procedure for testing the accuracy of the coordinates, association
datasets, naming, and classifications was to use the information
in the metadata files.  The HSLA SIMBAD, NED, or Phase II names
were input into the HST Mission Search form and were compared to
the returned datasets and programs with the program IDs listed in
the metadata file. The HST observation coordinates were compared
to the SIMBAD or NED object entry if it existed and investigated
if it seemed discrepant with the details and intended target of the
given HST programs. The vast majority of HSLA targets are isolated
objects with a well classified nature. Most classification and
naming failures occurred in the situation where a structure within
a nearby, spatially resolved galaxy was targeted and where SIMBAD
has multiple unique objects listed within 2$^{\prime\prime}$ of an
association's average coordinates. Users should use extra caution
when including a target association when multiple unique SIMBAD or
NED objects are present within 2$^{\prime\prime}$, which is flagged
in the metadata file for the target.

\begin{table}[h!]
\centering
\caption{Target association coordinate accuracy, target association
input dataset accuracy, target name accuracy, and target classification
accuracy. Limiting accuracies are calculated assuming binomial
probability distribution functions at 95\% confidence.}
\vspace{0.2cm}
\begin{tabular}{@{}lcc@{}}
\toprule
\textbf{HSLA Property} & \textbf{Accuracy} & \textbf{Limiting Accuracy} \\
\midrule
Coordinates & 99\% & 95\% \\
Association Inputs & 98\% & 94\% \\
Naming & 93\% & 87\% \\
Classifications & 92\% & 86\% \\
\bottomrule
\end{tabular}

\label{tab:randomtest}
\end{table}

Table \ref{tab:randomtest} provides our estimated accuracies for
association coordinates, association accuracy, naming accuracy, and
classification accuracy of the HSLA. For target association and
coordinate accuracies, we exceed our requirements, achieving 94\%
accuracy. For Classification and Naming, we find that we are better
than 86\% accurate.

\subsection{Flux and Wavelength Accuracy}

To ensure the flux and wavelength accuracy of our data products,
we perform a series of tests comparing each coadded spectrum with
its component xld datasets.  We use the analysis software developed
for HASP and described in the HASP ISR (Debes et al.\ 2024).  Briefly,
to test the flux accuracy, each xld spectrum is filtered to remove
pixels with serious data-quality flags, binned into 20 wavelength
bins, and compared with the similarly prepared coadded spectrum.
We consider only wavelength bins with S/N $> 20$ and only xld spectra
with at least five such wavelength bins.  For each wavelength bin,
we compute the residual $r = (F_{\rm xld} - F_{\rm coadd}) / F_{\rm
coadd}$.  For each xld spectrum, we seek to have both the mean and
standard deviation of the residuals be less than 0.05.  The wavelength
accuracy is assessed by cross-correlating the prepared (unbinned)
coadded and xld spectra of the same high-S/N spectra; our goal is
for any offset to be less than 1 pixel.

\begin{deluxetable}{l c c c c}
\tablewidth{0pt}
\tablecaption{HSLA Test Results \label{tab:HSLA_test_results}}
\tablehead{
\colhead{Grating} & \colhead{Total xld} & \colhead{Sufficient}  & \colhead{Flux}    & \colhead{Wavelength} \\
              & \colhead{Datasets}  & \colhead{Bins\tablenotemark{a}} & \colhead{Success} & \colhead{Success}}
\tablecolumns{5}
\startdata

\sidehead{COS}
G130M & 12966 & 71.45\% & 94.17\% & 98.46\% \\
G160M & 8652 & 67.83\% & 92.88\% & 99.76\% \\
G140L & 4832 & 18.63\% & 96.44\% & 84.00\% \\
G185M & 1462 & 59.78\% & 97.48\% & 97.14\% \\
G225M & 773 & 47.09\% & 98.35\% & 94.78\% \\
G285M & 312 & 92.31\% & 89.93\% & 89.58\% \\
G230L & 1408 & 58.95\% & 91.33\% & 89.88\% \\

\midrule

\sidehead{STIS}
G140M & 1098 & 37.16\% & 93.14\% & 99.02\% \\
G140L & 2391 & 64.83\% & 89.48\% & 99.03\% \\
G230L & 2267 & 65.06\% & 91.46\% & 99.73\% \\
G230LB & 679 & 91.75\% & 94.86\% & 97.11\% \\
G230M & 382 & 73.56\% & 94.31\% & 98.22\% \\
G230MB & 361 & 54.85\% & 99.49\% & 95.96\% \\
G430L & 1886 & 76.56\% & 92.31\% & 96.19\% \\
G430M & 960 & 61.88\% & 88.55\% & 91.75\% \\
G750L & 1539 & 89.86\% & 95.30\% & 97.90\% \\
G750M & 1011 & 65.38\% & 82.90\% & 87.90\% \\
E140H & 1083 & 96.77\% & 94.08\% & 99.05\% \\
E140M & 2401 & 89.09\% & 89.01\% & 96.35\% \\
E230H & 1643 & 95.86\% & 93.33\% & 99.49\% \\
E230M & 2377 & 90.66\% & 96.47\% & 99.44\% \\

\midrule

{\bf TOTAL}\tablenotemark{b} & 50483 & 67.2\% & 93.16\% & 97.54\% \\
\enddata
\tablenotetext{a}{Flux and wavelength tests were performed only on
xld files with at least five wavelength bins with S/N $> 20$.}
\tablenotetext{b}{Reported percentage values are averages.}
\end{deluxetable}

This analysis was performed for each of the xld spectra included
in the HSLA project.  Results are presented in Table
\ref{tab:HSLA_test_results}.  We see that the mean and standard
deviation of the residuals are less than 0.05 for more than 93\%
of xld spectra with S/N sufficient to make the comparison.  More
than 97\% of those spectra yield wavelength offsets of less than 1
pixel. Both the flux and wavelength tests substantially exceed the
benchmark success rate of 75\%.

\subsection{CALSPEC Standards and Flux Calibration}

All COS and STIS observations are flux calibrated using observations
of selected standard White Dwarf stars (WDs), for which high fidelity
synthetic model spectra are available in the CALSPEC database. The
detector responses on both instruments change with time, and in
order to maintain calibration accuracy, standard WDs are observed
regularly in order to characterize the time-dependent sensitivity
(TDS) for each of the modes. For most of the COS and STIS modes the
flux calibration accuracy is 5\% absolute and 2\% relative. In other
words, the fluxes reported in any calibrated spectrum is expected
to be within 5\% of the true value of the source flux, and for any
given observation, the relative fluxes across the bandpass within
a given mode and, in the case of COS, a given detector segment are
accurate to 2\%. The main exceptions, for which the flux calibrations
have larger uncertainties, are those COS modes with wavelength
coverage extending shortward of 1150\,\AA\ and the STIS high-resolution
echelle modes, especially those obtained using slits narrower than
0.2$^{\prime\prime}$. Additionally, for several modes the flux
calibration accuracies systematically degrade toward the detector
edges. (See the \href{https://hst-docs.stsci.edu/cosihb}{COS} and
\href{https://hst-docs.stsci.edu/stisihb}{STIS} Instrument Handbooks
and references therein for details.)

\begin{figure}
\centering
\includegraphics[width=1.0\textwidth]{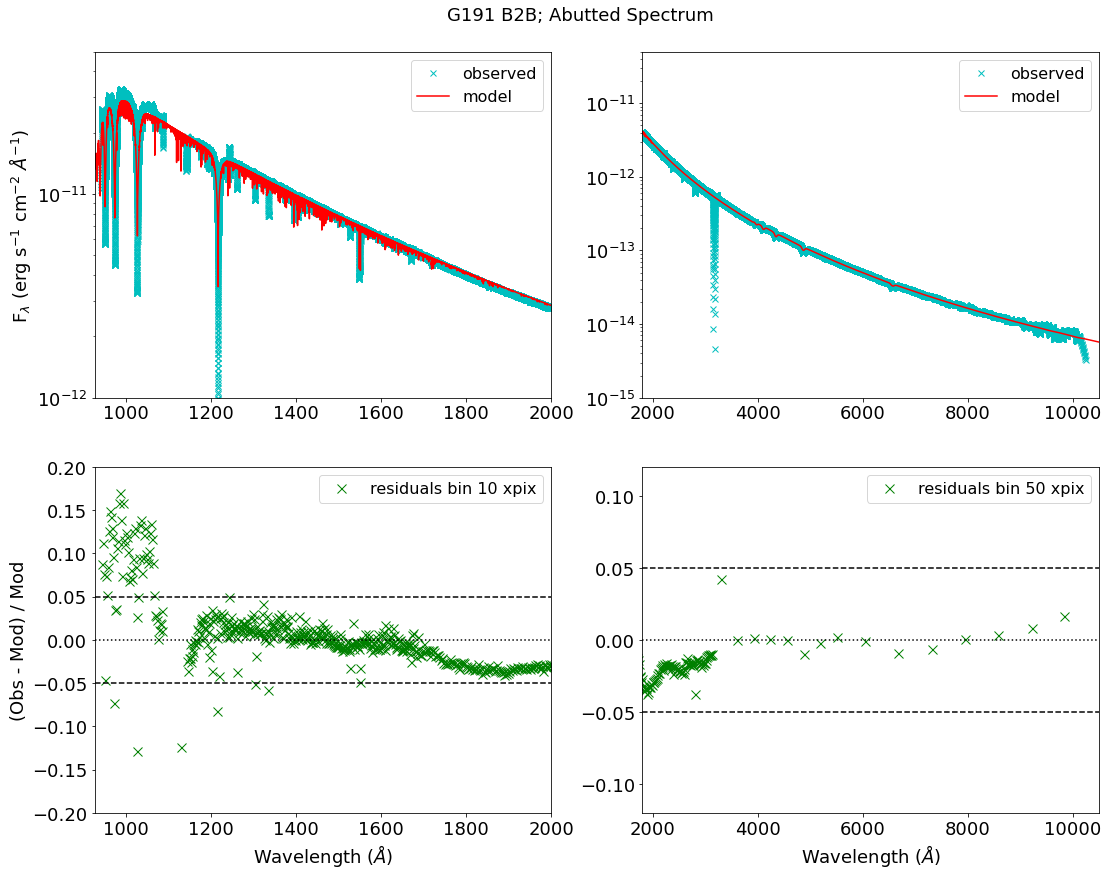}
\caption{The top panels show the HSLA abutted spectrum of G~191-B2B
overlaid on a synthetic spectrum of the star. The plot on the left
spans about 1100\,\AA\, and that on the right spans the remaining
$\approx$8000\,\AA\ covered by COS and STIS\@. The bottom panels
show residuals obtained after binning the observed and model spectra.
The bin sizes are different for the two sections of the spectrum,
and were chosen solely for clarity of display.}
\label{fig:aspec_g191b2b}
\end{figure}

\begin{figure}
\centering
\includegraphics[width=1.0\textwidth]{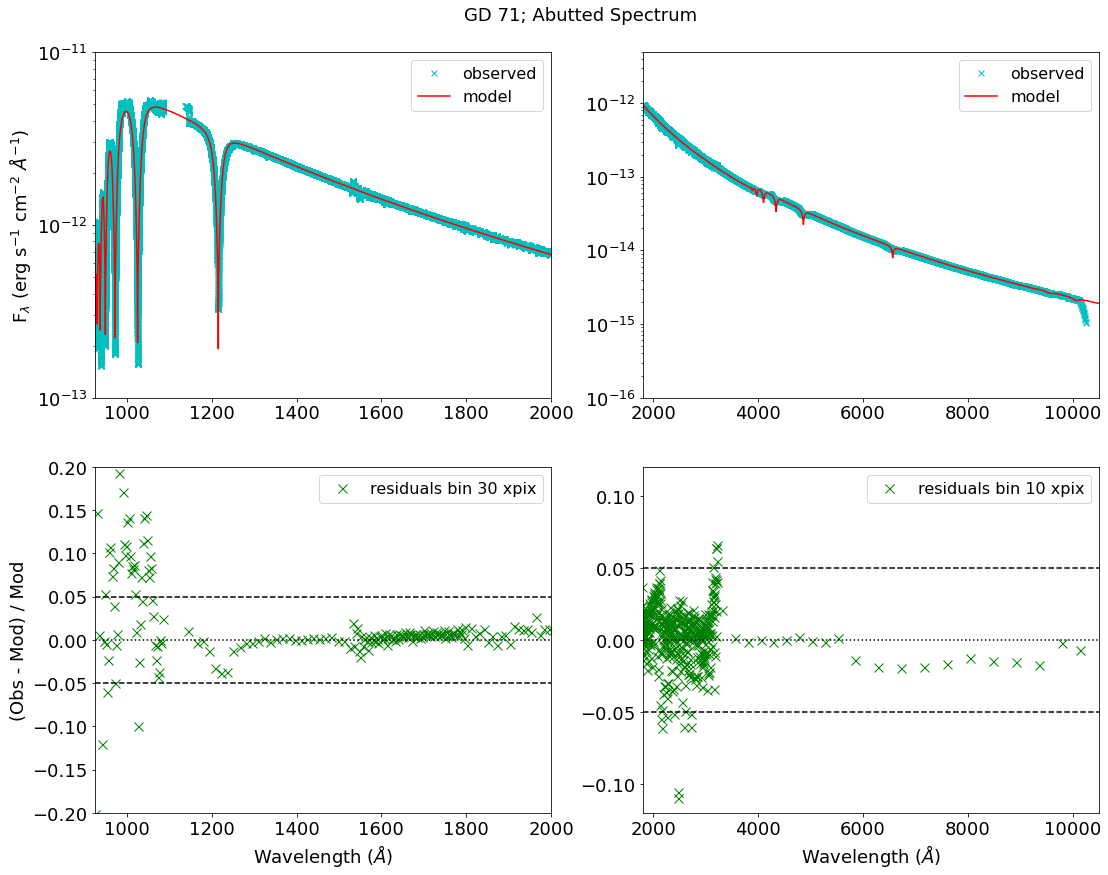}
\caption{Same as Fig.~\protect\ref{fig:aspec_g191b2b} but for GD~71.}
\label{fig:aspec_gd71}
\end{figure}

\begin{figure}
\centering
\includegraphics[width=1.0\textwidth]{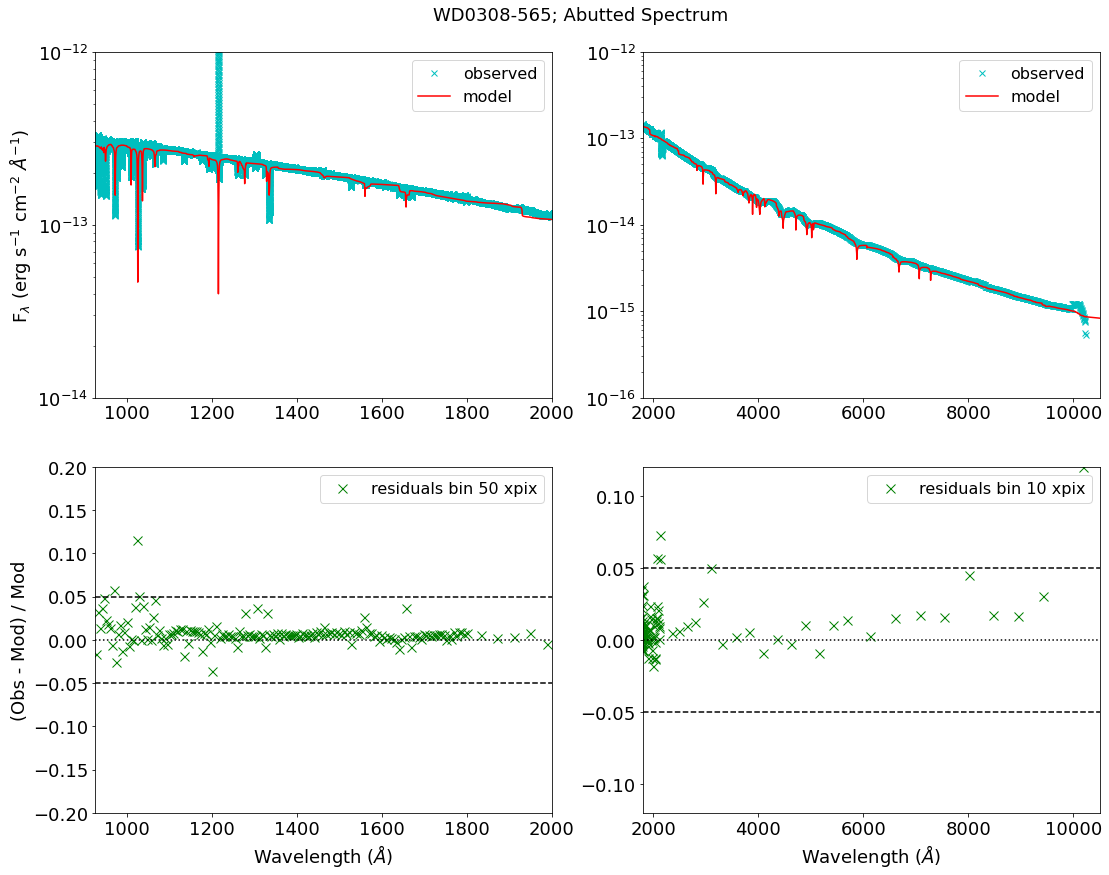}
\caption{Same as Fig.~\protect\ref{fig:aspec_g191b2b} but for WD0308-565.}
\label{fig:aspec_wd0308}
\end{figure}

The flux calibration accuracy of the HSLA coadded spectra is limited
by the accuracy of the calibrated input data. Additional sources
of uncertainty are present in the coadds, since they typically
include spectra obtained with several modes, and taken over longer
spans of time than individual observations. In order to characterize
the flux calibration accuracy of the HSLA data products we examine
the residuals between coadded spectra of standard WDs and their
synthetic models. We focus on three standards, G~191-B2B, GD~71 and
WD0308-565, which together span all the COS and STIS modes used,
and which have been observed as part of numerous programs either
for flux calibration or for monitoring the TDS\@. The HSLA abutted
spectra (stored in the ``aspec.fits'' files) for each of these stars
overlaid on the correspending models are shown in
Figs.~\ref{fig:aspec_g191b2b}--\ref{fig:aspec_wd0308}. Also shown
are the residuals between binned data and model spectra.  Note that
the bin sizes are not uniform across the entire wavelength range
for abutted spectra (\S4) and we have chosen a binning value suitable
for display in each case.

The overlays show that the data and models generally agree well
across the spectrum.  One exception that stands out is the short
wavelength region ($<$1100\,\AA) for G~191-B2B
(Fig.~\ref{fig:aspec_g191b2b}).  This is a case where there are
only a handful of observations using the COS modes that cover this
region, and it is known that the model does not provide a good fit
at these wavelengths. More subtle issues are seen in the case of
GD~71 at the shortest wavelengths where the spectrum is dominated
by deep hydrogen lines, and where the model may not be sufficiently
accurate. The other small discrepancy seen is a systematic offset
of about 2\% longward of around 6000~\AA\ for both GD~71 and
WD0308-565.  The abutted spectra are meant as quicklook data products
and not suitable for quantitative tests of the flux calibration
accuracy. For that we turn to the scientifically usable spectra
that form the core data products provided by HSLA\@.

For each of the three standard stars we compare the spectra in
individual ``cspec.fits'' files against the models. These spectra
are grating-dependent, and in the case of COS FUV spectra HSLA
separated by LP (\S4). The COS NUV detector has operated
at only a single lifetime position, which is designated LP1, and STIS
does not use lifetime positions. For testing the HSLA flux calibration
accuracy we restrict ourselves to STIS data for G~191-B2B and to COS
data for WD0308-565, while for GD~71 we include both COS and STIS modes.

To compare models with data, we first obtain the residuals at the
full spectral resolution of the HSLA spectra (i.e.\ without binning):

\begin{flushleft} 
$Residual = (F_{HSLA} - F_{model})/F_{model}$
\end{flushleft}

\noindent and then we use the routine ``astropy.stats.sigma\_clip''
to obtain the means and standard deviations of the residuals,
rejecting 3$\sigma$ outliers iteratively for a maximum of five
iterations. The resulting means and standard deviations are used
as measures of flux calibration accuracy for each mode.

These results are shown in
Tables~\ref{tab:g191b2b_residuals}--\ref{tab:wd0308_residuals} for
the three stars. In each table, the first column is the grating
(and for COS modes the LP), the second and third columns give the
wavelength range of the spectra, the fourth column is the total
number of wavelength points, the fifth column is the number of these
points omitted due to sigma-clipping, and the sixth and seventh
columns are the residual means and standard deviations.

G~191-B2B is one of the main calibrators for STIS. It has been
observed using all STIS modes except for the low-resolution MAMA
modes, G140L and G230L. The residual mean is less than 1\% for all
the modes except the two echelle modes E230H and E230M where there
are offsets of about $-$2\% and $-1$\%, respectively
(Table~\ref{tab:g191b2b_residuals}). The standard deviations range
from 0.4\% (G230LB, G430L) to around 3.5\% (G140M and G750M). The
star has been observed using COS, but those data are known to have
issues with flux calibrations and are not included in the table.

GD~71 has been used for calibrating the COS NUV channels, and for
tracking the TDS of COS FUV grating G160M, segment A. It is also
used to calibrate or verify several STIS MAMA and CCD modes.
Table~\ref{tab:gd71_residuals} shows that the means are mostly less
than 1\%, and in the case of G160M at LP1 and G750L between 1 and
2\%. The one exception is G130M at LP2, where the residual mean is
6\% and standard deviation 8\%. This is because the only G130M
CENWAVEs used to observe GD~71 have been c1055 and c1096 and using
only detector segment B, as reflected by the minimum and maximum
wavelengths listed. Furthermore, there has been only one observation
using c1055, where the wavelength extends below the Lyman cut-off
at 912\,\AA\@. The flux calibration between $\approx$912 and 940\,\AA\
is poor and that region has been omitted in our analysis.

WD0308-565 is the primary calibrator for the COS FUV modes, and
is also used to monitor the TDS for most of them. The residual means
(Table~\ref{tab:wd0308_residuals}) are all less than 1\% except for
G130M at LP2 where it is 1.7\%. The standard deviations fall in a
narrow range between 2.2 and 3.5\%. Note that in this case, as
WD0308-565 has been observed many times using CENWAVE 1055 for TDS
monitoring, we have retained the default minimum wavelength for
G130M at LP2. This results in the relatively large number of clipped
points.

Across all the targets and modes, most of the residual means are
less than 1\%, and the residual standard deviations have a median
of 2.3\%. Based on these results we conclude that the HSLA ``cspec''
spectra retain the absolute flux calibration accuracy of 5\% of
individual observations and have a relative accuracy of 2.3\%, which
is only slightly lower than for individual observations.

\begin{deluxetable}{lcccccc}
\tablewidth{0pt}
\tablecaption{HSLA Flux Calibration: G~191-B2B\label{tab:g191b2b_residuals}}
\tablehead{
\colhead{Grating}
& \colhead{$\lambda_{min}$}  
& \colhead{$\lambda_{max}$}    
& \colhead{N$_{tot}$} 
& \colhead{N$_{clipped}$}
& \colhead{Residual}
& \colhead{Residual} \\

\colhead{} & \colhead{(\AA)} & \colhead{(\AA)}
& \colhead{} & \colhead{} & \colhead{Mean} & \colhead{St. Dev.}
}
\tablecolumns{7}
\startdata
\sidehead{STIS ECHELLE}
E140H &   1141.0 &  1687.9 &   71933 &     5397 & -0.000 &  0.023 \\
E230H &   1629.4 &  3146.3 &  111659 &     2388 & -0.021 &  0.024 \\
E140M &   1140.4 &  1709.7 &   30545 &     2955 &  0.004 &  0.020 \\
E230M &   1607.1 &  3073.5 &   29138 &     1137 & -0.012 &  0.015 \\
\sidehead{STIS MAMA}
G140M &   1145.3 &  1741.5 &   11138 &      852 &  0.007 &  0.035 \\
G230M &   1642.3 &  3097.1 &   16438 &     1161 & -0.001 &  0.032 \\
\sidehead{STIS CCD}
G230MB &   1635.1 &  3184.7 &   10223 &     1099 &  0.001 &  0.017 \\
G430M &   3022.1 &  5610.1 &    9300 &      246 & -0.001 &  0.015 \\
G750M &   5450.6 & 10127.5 &    8401 &      999 & -0.002 &  0.035 \\
G230LB &   1664.4 &  3073.4 &    1026 &       53 & -0.001 &  0.004 \\
G430L &   2965.5 &  5709.8 &     999 &       11 & -0.001 &  0.004 \\
G750L &   5294.9 & 10258.1 &    1017 &      211 & -0.005 &  0.017 \\
\enddata
\end{deluxetable}

\begin{deluxetable}{lcccccc}
\tablewidth{0pt}
\tablecaption{HSLA Flux Calibration: GD~71\label{tab:gd71_residuals}}
\tablehead{
\colhead{Grating / LP\tablenotemark{a}} 
& \colhead{$\lambda_{min}$}  
& \colhead{$\lambda_{max}$}    
& \colhead{N$_{tot}$} 
& \colhead{N$_{clipped}$}
& \colhead{Residual}
& \colhead{Residual} \\

\colhead{} & \colhead{(\AA)} & \colhead{(\AA)}
& \colhead{} & \colhead{} & \colhead{Mean} & \colhead{St. Dev.}
}
\tablecolumns{7}
\startdata
\sidehead{COS FUV}
G130M / 2\tablenotemark{b}   & 940.0 &  1080.0 & 14103 & 377 &  0.062 &  0.081 \\
G160M / 1 & 1602.1 &  1771.7 &   13858 &      259 &  0.011 &  0.086 \\
G160M / 2 & 1575.0 &  1801.5 &   18494 &      595 &  0.006 &  0.023 \\
G160M / 3 & 1575.0 &  1801.6 &   18495 &      744 &  0.001 &  0.024 \\
G160M / 4 & 1533.3 &  1800.9 &   21839 &      767 &  0.003 &  0.023 \\
G160M / 6 & 1530.7 &  1801.1 &   22070 &      835 &  0.007 &  0.023 \\
\sidehead{COS NUV}
G185M / 1 & 1662.1 &  2135.4 &   12442 &      242 &  0.013 &  0.049 \\
G225M / 1 & 2068.4 &  2526.4 &   12636 &      232 & -0.002 &  0.043 \\
G285M / 1 & 2473.8 &  3226.9 &   17303 &      340 &  0.001 &  0.057 \\
G230L / 1 & 1700.4 &  3099.8 &    3373 &      224 &  0.003 &  0.030 \\
\sidehead{STIS MAMA}
G140M & 1145.5 &  1741.3 &   11117 &      205 & -0.003 &  0.024 \\
G230M & 1724.6 &  3096.8 &   15518 &      209 &  0.002 &  0.023 \\
G140L & 1137.9 &  1718.0 &     994 &      102 & -0.004 &  0.007 \\
G230L & 1579.4 &  3168.1 &    1024 &       34 &  0.001 &  0.003 \\
\sidehead{STIS CCD}
G230LB & 1663.4 &  3073.8 &    1027 &       41 & -0.003 &  0.003 \\
G430L  & 2969.5 &  5708.3 &     997 &       25 &  0.000 &  0.004 \\
G750L  & 5302.9 & 10256.4 &    1015 &      178 & -0.018 &  0.021 \\
\enddata
\tablenotetext{a}{Lifetime Positions relevant only for COS FUV modes.}
\tablenotetext{b}{The minimum wavelength has been set to exclude G130M/1055-only data (see text).}
\end{deluxetable}

\begin{deluxetable}{lcccccc}
\tablewidth{0pt}
\tablecaption{HSLA Flux Calibration: WD0308-565\label{tab:wd0308_residuals}}
\tablehead{
\colhead{Grating / LP}  
& \colhead{$\lambda_{min}$}  
& \colhead{$\lambda_{max}$}    
& \colhead{N$_{tot}$} 
& \colhead{N$_{clipped}$}
& \colhead{Residual}
& \colhead{Residual} \\

\colhead{} & \colhead{(\AA)} & \colhead{(\AA)}
& \colhead{} & \colhead{} & \colhead{Mean} & \colhead{St. Dev.}
}
\tablecolumns{7}
\startdata

\sidehead{COS FUV}
G130M / 1 & 1132.7 & 1472.1 &   34037 &     2259 &  0.007 &  0.028 \\
G130M / 2 &  900.4 & 1472.3 &   57343 &    11173 &  0.006 &  0.029 \\
G130M / 3 & 1066.4 & 1472.2 &   40678 &     4477 &  0.002 &  0.023 \\
G130M / 4 & 1065.1 & 1472.2 &   40814 &     3904 &  0.005 &  0.022 \\
G130M / 5 & 1131.3 & 1472.5 &   34202 &     3604 &  0.017 &  0.028 \\
\\
G160M / 2 & 1387.6 & 1795.9 &   33343 &      714 &  0.008 &  0.035 \\
G160M / 3 & 1384.0 & 1802.0 &   34123 &     1169 &  0.000 &  0.025 \\
G160M / 4 & 1343.1 & 1802.0 &   37446 &     1711 &  0.008 &  0.023 \\
G160M / 6 & 1339.6 & 1801.8 &   37722 &     1490 &  0.007 &  0.027 \\
\\
G140L / 2 &  915.0 & 1959.9 &   12962 &     1250 &  0.004 &  0.034 \\
G140L / 3 &  915.0 & 1959.9 &   12962 &     1290 & -0.001 &  0.022 \\
G140L / 4 &  915.0 & 1959.9 &   12963 &     1075 &  0.004 &  0.027 \\
\enddata
\end{deluxetable}

\vspace{-0.3cm}
\ssection{Accessing the Catalog and Data}\label{sec:accessdata}

Users can access HSLA data through the
\href{https://mast.stsci.edu/portal/Mashup/Clients/Mast/Portal.html}{MAST
Portal}, the \href{https://mast.stsci.edu/search/ui/#/hst}{HST
Mission Search Form} or via \texttt{astroquery}.  In addition to
the usual search categories such as target name and coordinates,
HSLA products may be found using the classifications in any of the
three tiers.  Data may be obtained for individual targets, or for
all targets spanning one or more classes. Metadata files are included
in the data products and may be downloaded separately if necessary.

Data access may change with time and users are encouraged to visit
the \href{https://archive.stsci.edu/missions-and-data/hst/hsla}{HSLA
website} for the latest details on how to access data.

\subsection{Considerations for HSLA data products, associations, and classifications}

While HSLA data products, associations, and classifications are
robust in the large majority of cases, certain target types or
observation types are not well suited for coadds across multiple
programs. We provide a brief list of considerations here, and users
are encouraged to carefully inspect the trailer files, metadata
files, and data before relying on the results.

\begin{itemize}

\item{In crowded fields, particularly where there are both multiple
HST pointings and SIMBAD objects, it will be useful to check the
individual program target specifications and information to determine
whether the coadds and classifications are correct.  If the HSLA
data are compromised by including multiple targets, then HASP data
may be used or a custom coaddition of user selected datasets.}

\item{Blue Modes of G130M have differing resolutions and thus
differing LSFs; caution should be used when analyzing lines in
overlap regions with the 1055,1096, 1222, and 1291 CENWAVES. An
example HSLA notebook of handling differing LSFs is given in Section
\ref{sec:custom}.}

\item{There is a small wavelength offset for spectra taken at the
E1 position that is currently unaccounted for in the STIS pipeline
and needs to be corrected for (See Section 4.2 of Joyce et al.,
2018). This can be mitigated by applying a wavelength correction
and performing a custom coaddition.}

\item{The HSLA code does not implement wavelength shifts to the
input files in the sense of matching up features in spectra to
improve the wavelength calibration. The data are resampled
onto a new grid and so wavelengths can shift by up to half a
wavelength bin. Radial velocity variability of sources will not
be accounted for or corrected.}

\item{The flux accuracy for some modes like the STIS echelles,
narrow slits on STIS (below 0.2") or the COS 1055 and 1096 CENWAVES
can be less precise than our flux requirements either due to absolute
flux calibration accuracy or focus variations within an HST orbit.}

\item{Target associations, target names, and target classifications
are only as reliable as their SIMBAD entries for a majority of HSLA
targets. Users who find an inaccuracy are encouraged to submit a
help desk ticket so that corrections to the metadata files can be
made as well as to SIMBAD.}

\item{Despite a high degree of accuracy between the target associations
and SIMBAD classifications, completeness for any object depends on
details of the classifications. No effort was made to be consistent
for e.g. objects with multiple properties such as post-AGB stars
vs. Planetary nebulae central stars or Active galaxies that also
are a particular morphology. In this case cross matching with a
known catalog of objects may be more complete.}

\item{If multiple Phase II classifications were present within a
program or across different programs the most common property was
used, with some limited filtering of special cases.}

\item{Relatively compact sources that are spatially resolved with
STIS and COS often meet our requirements but may suffer from excessive
dataset filtering or flux mismatches between gratings. Users should
carefully understand the different angular areas subtended by each
dataset to properly interpret the HSLA coadds.}

\item{Sources that are variable in time are likely to have excessive
dataset filtering or flux mismatches between gratings. Users should
carefully understand the different epochs in which HST observations
were taken to properly interpret the HSLA coadds for a variable
source. Custom coaddition of datasets scaled to a common flux may
be appropriate for some science cases. An example of how to do this
is given in the HASP notebook on
\href{https://spacetelescope.github.io/hst_notebooks/notebooks/HASP/FluxScaleTutorial/FluxScaleTutorial.html}{flux
scaling}.}

\item{Moving Targets are not included in HSLA under the assumption
that they are often flux variable.}

\end{itemize}

\ssection{HSLA Examples} \label{subsec:dpexamples}

In this Section we give an introduction to some of the potential
unique uses of HSLA data products. HSLA products are broadly useful
in that they can facilitate the analysis of HST spectra for objects
that have been observed many times or with many modes. Collating
multiple programs then increases overall SNR, and extends wavelength
coverage. Additionally, target classification enables studies on
classes of objects, such as stars or galaxies, or subclasses of
objects such as Seyfert 1 galaxies or WDs.

\subsection{High SNR Spectrum of WD0308-565 from 920-10000\AA}
\begin{figure}
\centering
\includegraphics[width=0.9\textwidth]{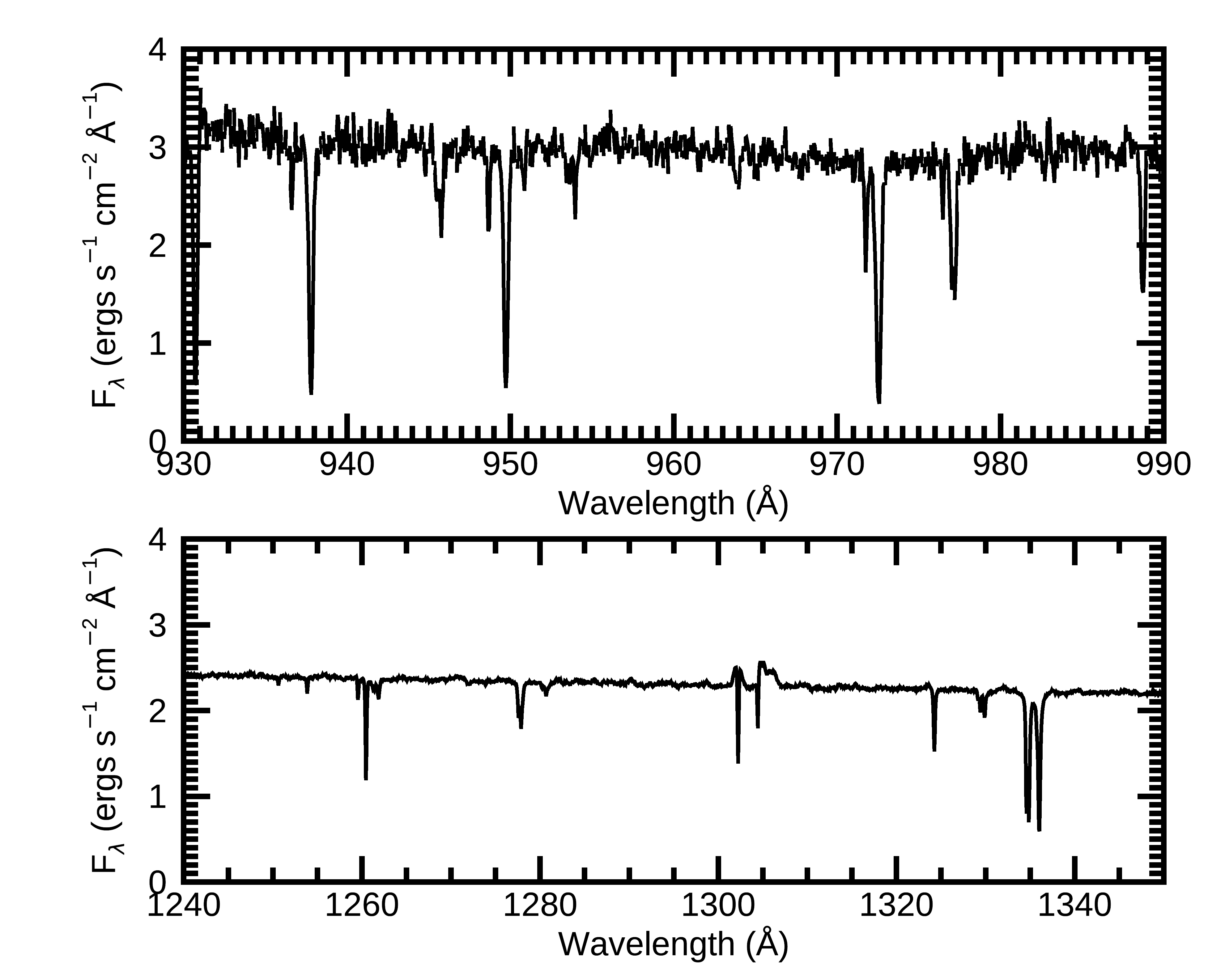}
\caption{Selected regions of the HSLA quick look spectrum of
WD0380-565, which spans from 920-10,000~\AA. Top: Spectrum near
950~\AA\ that shows several ISM H lines. Bottom: Several ISM Si and
Photospheric C lines are seen in a very high SNR region of the FUV
spectrum, obtained by dozens of individual observations of this
flux standard. Residual OI airglow lines are present around 1301
and 1306~\AA.}
\label{fig:wd0308}
\end{figure}

WD0308-565 is a DB white dwarf that has been extensively used as a
calibrator for the COS instrument. In the HSLA, its SIMBAD name is
WG~7 and its unique target identifier is 722. Because of its status
as a calibrator, many COS modes have been observed throughout the
operational lifetime of the instrument (\S5.3). As an example, we
show portions of the 920-10,000~\AA\ spectrum in Figure \ref{fig:wd0308}
that is comprised of a mix of medium resolution COS spectra and low
resolution STIS spectra from six different gratings.

In particular, the median effective exposure time in the region of
1200-2000~\AA\ is 1.1$\times$10$^5$~s, or 30.8 hours of exposure
time. At this total exposure time, one is limited by fixed pattern
noise. Even so, the fixed pattern noise is diminished due to the
combination of several lifetime positions, FP-POS and small
non-repeatable shifts of the OSM1 mechanism. We measure an SNR of
111 in the region of 1220-1240~\AA\ assuming a $\sim$6 bin resolution
element, which exceeds the nominal best SNR for G130M mode assuming
COS 2025 rules by a factor of nearly 4.

Below 1000~\AA, the total SNR is close to 36/resolution element,
allowing studies of the ISM and the carbon abundance of this DB in
exquisite detail, with the caveat of a more complex line spread
function due to multiple lifetime positions. If a science case
requires an accurate measure of the line spread function, users can
refer to the HSLA notebook that describes how to create a custom
LSF that accounts for the merging of different lifetime positions,
or they can make use of the lifetime position single grating spectra.

\subsection{UV Coverage of Exoplanet Host Star $\beta$~Pictoris}

\begin{figure}
\centering
\includegraphics[width=0.9\textwidth]{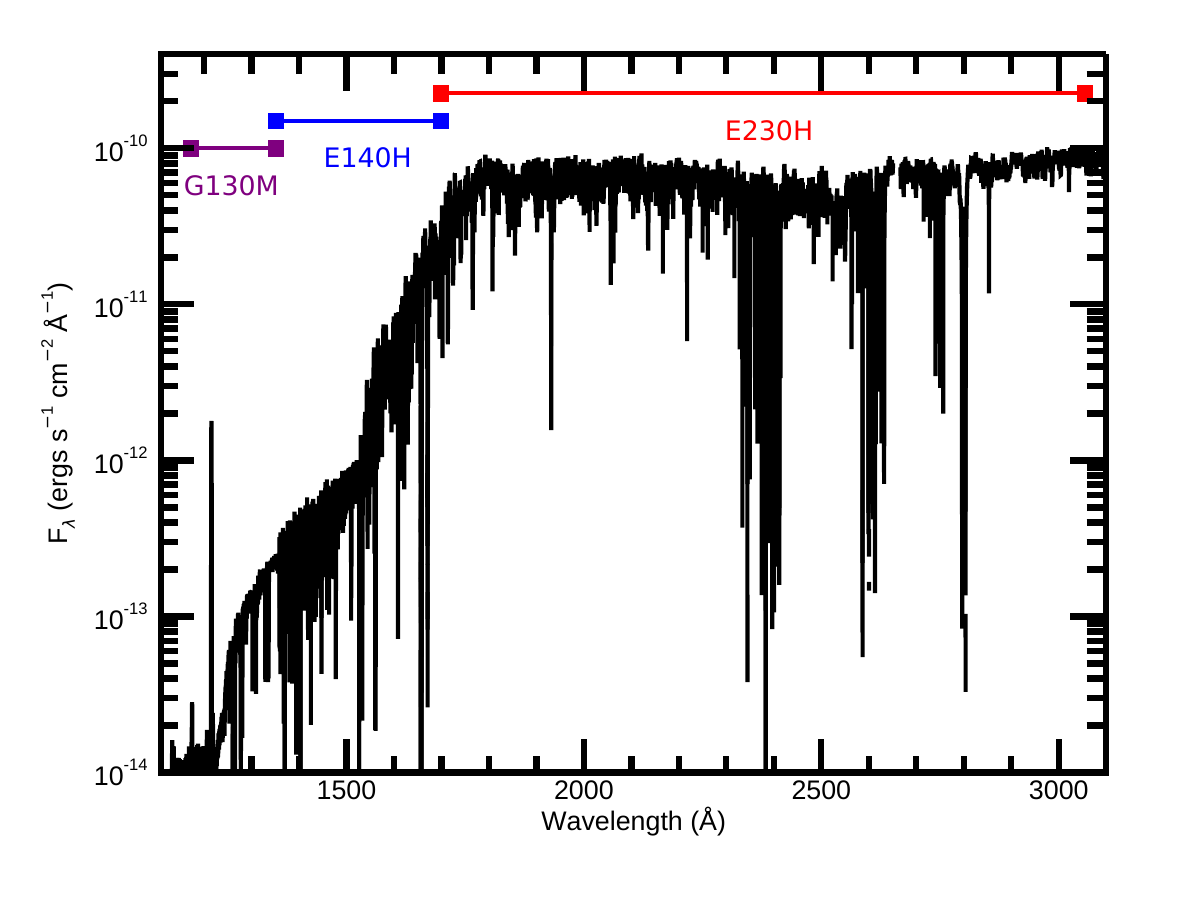}
\caption{Full UV SED of $\beta$~Pictoris derived from the HSLA
quicklook \texttt{aspec} file, covering three separate COS and STIS
gratings. Abundant stellar and circumstellar absorption features
are seen in the spectrum.}
\label{fig:betapic}
\end{figure}

Eleven separate programs (PIDs: 7512, 9154, 13406, 14089, 14735,
14894, 15396, 15412,15479, 17421, 17790) have targeted the well-known
exo-comet, giant-planet, and debris disk hosting star in the UV
using a wealth of COS and STIS modes, including G130M, E140H, and
E230H. For HSLA, the target name is ``* Beta Pic'' and its unique
target identifier is 1134. In addition to the photosphere, the
spectrum includes absorption due to a non-varying stable gas component
roughly at the rest velocity of the star and time variable red- and
blue-shifted components at high velocities caused by the evaporation
of exo-comets. Figure \ref{fig:betapic} shows the full 1135-3095~\AA\
spectrum that spans four orders of magnitude in flux for this
interesting star, where multiple elements in the circumstellar gas
disk have been detected (Roberge et al., 2006; Wilson et al., 2017;
Wilson et al., 2019)

One caveat to note is that since time variable absorption occurs
at higher redshift and blueshift from the system radial velocity,
the quicklook spectrum shown here might not be desirable for
understanding the short term evaporation behavior of $\beta$~Pictoris'
exo-comets. In that case, users are recommended to investigate the
visit and program level HASP coadds for the individual programs
listed above. Inspection of the HSLA metadata file would facilitate
obtaining that information quickly for any HSLA object to check for
variability.

\subsection{Large scale UV spectroscopic surveys of White Dwarfs}

Isolated WDs that are not actively accreting material have very
simple atmospheres that usually contain either hydrogen or helium
spectral lines. To that end, especially in the UV, they represent
an interesting population to observe in order to understand the
composition and properties of the local ISM. The WDs act as a
background light source on which ISM absoprtion is imprinted.
Accreting white dwarfs that show Si are primarily believed to be
accreting rocky material consistent with bulk Earth and thus are
probing the minor body populations of planetary systems that have
survived post-main sequence evolution (Zuckerman et al., 2007;
Koester et al., 2014).

While a detailed study of the ISM or WD accretion is beyond the
scope of this ISR, we demonstrate the potential power of HSLA in
facilitating large scale spectroscopic studies in the UV of different
types of astrophysical objects. In this example, we show that one
can easily generate a large survey of WDs observed by HST by
performing equivalent width measurements in the Si~1260~\AA\ FUV
ISM line  and the Si~1265~\AA\ photospheric accretion line resolved
and simultaneously covered by COS G130M, G140M, E140M, and E140H.

\begin{figure}
\centering
\includegraphics[width=0.9\textwidth]{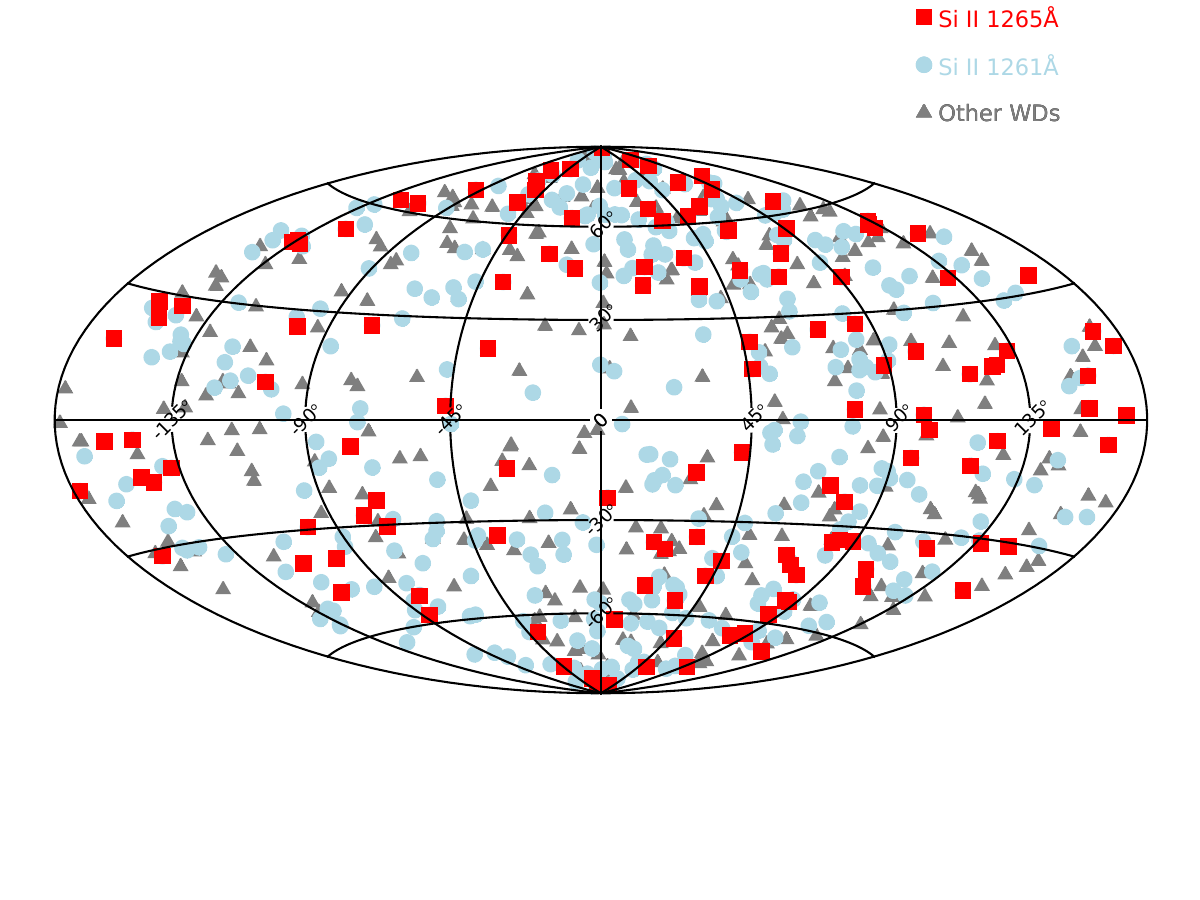}
\caption{An aitoff projection of galactic coordinates for HSLA white
dwarfs. The grey triangles are WDs that have sufficient wavelength
coverage and resolution to resolve and detect the FUV Si lines at
1261~\AA\ and 1265~\AA. Light blue circles represent HSLA WDs with
a significant Si 1261~\AA\ line, indicative of detectable local ISM
absorption, while red squares represent HSLA WDs with a significant
Si 1265~\AA\ line, indicative of ongoing accretion onto the WD from
rocky material.}
\label{fig:wd}
\end{figure}

First, we select all HSLA targets that are classified as "White
Dwarfs", which results in 890 HSLA targets. We then read each aspec
file to select a subset of HSLA targets that have wavelength coverage
around 1260~\AA. Roughly 753 pass that selection. After doing a
simple fit to the contiuum, we calculate the equivalent width within
$\pm$50~km~s$^{-1}$ of the rest line center to get a rough estimate
of the absorption due to the ISM for each target. We also measure
the equivalent width of the Si~1265~\AA\ line, which is generally
only detected if there is active accretion onto the photosphere of
the WD (Koester et al., 2014).  We required that a spectrum have
at least two wavelength bins in our equivalent width measurement
region, which ruled out the lower resolution modes. We assume each
WD with just a significant 1261~\AA\ detection is a clean ISM
detection and each WD with a significant 1265~\AA\ detection is
appropriate for studying WD accretion. We assume any equivalent
width measurement larger than 3~$\sigma$ to be significant, using
the nearby continuum to estimate our uncertainties in equivalent
width. From this sample, we find 462 WDs with detectable ISM accretion
and 135 WDs with detectable Si from the accretion of rocky material.
Figure \ref{fig:wd} shows the distribution of the HSLA WD targets
in galactic coordinates.

\vspace{-0.3cm}
\ssection{Jupyter Notebooks for Custom Coadds}\label{sec:custom}

The coadd code and its HASP/HSLA wrapper are available for public
use in Space Telescope's GitHub Organization; \texttt{coadd} is
located in the
\href{https://github.com/spacetelescope/ullyses/}{ULLYSESrepository}, and
the HASP/HSLA wrapper for \texttt{coadd} is in the
\href{https://github.com/spacetelescope/hasp}{HASP repository}.
Both codes are scripted in Python (compatible with versions 3.9 or
later) and are installable using pip. More detailed installation
and use instructions are given in our tutorial notebooks, which are
described in the HASP ISR (Debes et al., 2024).  Two additional
notebooks specifically address the HSLA data products and the
challenges inherent in combining data from observing programs that
are separated in time by many years.

The first notebook,
\href{https://github.com/spacetelescope/hst_notebooks/tree/main/notebooks/HSLA/Intro/Intro.ipynb}
{``Introduction to the HSLA Data Products and Tools''}, explores
the standard HSLA data files returned by MAST.  It provides a simple
example of a custom coadd, demonstrating how to adjust the logic
by which the abutting routine decides which coadded spectra (cspec
files) to use for each wavelength region.

The second notebook,
\href{https://github.com/spacetelescope/hst_notebooks/tree/main/notebooks/HSLA/Setup/Setup.ipynb}
{``Combining COS Data from Multiple Lifetime Positions and Central
Wavelengths''}, explores how the COS line-spread function (LSF)
varies with LP and CENWAVE.  Because the HASP/HSLA script sums all
of the spectra from a particular grating without regard to LP or
CENWAVE, its final products may not have the highest-possible
spectral resolution.  The notebook presents several techniques for
dealing with these effects.

All HASP and HSLA notebooks are available from the
\href{https://spacetelescope.github.io/hst_notebooks}{Space Telescope
HST Notebook Repository HQ.}

\ssection{Conclusions} \label{sec:conclusion}

This report introduces the methods and advantages of a new Hubble
Spectroscopic Legacy Archive. The new HSLA represents a critical
evolution in delivering spectroscopic data that leverages 28 years
of observations from STIS and 16 years of observations from COS,
allowing the study of objects with higher SNR spectra or greater
wavelength coverage than the data from individual programs. The
combination of coadded spectra associated with astrophysical object
types allows the study of galaxies and stars as large populations
that enhances the legacy value of HST spectroscopy for years to
come. Finally, the HSLA innovates a new way to classify and associate
datasets that could be applicable to other observatories that host
spectroscopic data products.



\vspace{-0.3cm}
\ssectionstar{Change History for COS ISR 2025-18}\label{sec:history}
\vspace{-0.3cm}
Version 1: 6 November 2025 - Original Document 

\vspace{-0.3cm}
\ssectionstar{References}\label{sec:references}
\vspace{-0.3cm}

\noindent
Debes, J., et al.\ 2024 COS Instrument Science Report 2024-01(v2) \\
Frey, K., \& Accomazzi, A. 2018, ApJS, 236, 24F \\ 
Joyce, S.~R.~G., Barstow, M.~A., Holberg, J.~B., et al.\ 2018, MNRAS, 481, 2361 \\
Koester, D., G{\"a}nsicke, B.~T., \& Farihi, J.\ 2014, A\&a, 566, A34 \\
Peeples, M. et al., 2017, COS Instrument Science Report 2017-4 \\
Roberge, A., Feldman, P.~D., Weinberger, A.~J., et al.\ 2006, Nature, 441, 724 \\
Roman-Duval, J., et al.\ 2025 ApJ, 985, 109 \\
Wilson, P.~A., Kerr, R., Lecavelier des Etangs, A., et al.\ 2019, A\&A, 621, A121 \\
Wilson, P.~A., Lecavelier des Etangs, A., Vidal-Madjar, A., et al.\ 2017, A\&A, 599, A75 \\
Zuckerman, B., Koester, D., Melis, C., et al.\ 2007, ApJ, 671, 872 \\


\end{document}